\documentclass[10pt,reqno]{amsart}
\usepackage[latin1]{inputenc}
\usepackage{xcolor}
\usepackage{url}
\usepackage{enumitem}
\usepackage{amstext}
\usepackage{amsthm}
\usepackage{amssymb}
\usepackage{graphicx}
\PassOptionsToPackage{normalem}{ulem}
\usepackage{ulem}
\usepackage[unicode=true,pdfusetitle,
 bookmarks=true,bookmarksnumbered=false,bookmarksopen=false,
 breaklinks=false,pdfborder={0 0 0},pdfborderstyle={},backref=false,colorlinks=true]
 {hyperref}

\makeatletter

\providecommand{\tabularnewline}{\\}
\providecolor{lyxadded}{rgb}{0,0,1}
\providecolor{lyxdeleted}{rgb}{1,0,0}

\DeclareRobustCommand{\lyxsout}[1]{\ifx\\#1\else\sout{#1}\fi}

\numberwithin{equation}{section}
\numberwithin{figure}{section}
\theoremstyle{plain}
\newtheorem{thm}{\protect\theoremname}
\theoremstyle{remark}
\newtheorem{rem}[thm]{\protect\remarkname}
\theoremstyle{definition}
\newtheorem{defn}[thm]{\protect\definitionname}
\theoremstyle{definition}
\newtheorem{example}[thm]{\protect\examplename}

\@ifundefined{date}{}{\date{}}
\usepackage{pdfsync}
\usepackage{graphics}
\usepackage{epstopdf}
\usepackage[all]{xy}
\usepackage{amscd}
\setlist[enumerate]{leftmargin=*,label=(\roman*),align=left}
\setlist[itemize]{leftmargin=*,align=left}

\newcommand{\xyR}[1]{ \makeatletter
\xydef@\xymatrixrowsep@{#1} \makeatother} 
\newcommand{\xyC}[1]{ \makeatletter
\xydef@\xymatrixcolsep@{#1} \makeatother} 
\entrymodifiers={++[ ][F]} 

\newcommand{\ra}{\longrightarrow}
\newcommand{\xra}[1]{\xrightarrow{\ \ #1\ \ }} 

\newcommand{\field}[1]{\mathbb{#1}}
\newcommand{\R}{\field{R}} 
\newcommand{\N}{\field{N}} 

\newcommand{\eps}{\varepsilon} 
\renewcommand{\phi}{\varphi}
\newcommand{\diff}[1]{\,\hbox{\rm d}#1} 

\DeclareMathOperator{\act}{ac}
\DeclareMathOperator{\stateNeigh}{n}

\newcommand{\tst}{t_{\rm st}}
\newcommand{\tend}{t_{\rm end}}

\newcommand{\ag}[1]{{\rm ag}(#1)}
\newcommand{\pa}[1]{{\rm pa}(#1)}
\newcommand{\pr}[1]{{\rm pr}(#1)}
\newcommand{\tist}[1]{t_{#1}^{\rm s}}
\newcommand{\tong}[1]{t_{#1}^{\rm o}}
\newcommand{\tarr}[1]{t_{#1}^{\rm a}}
\newcommand{\tfirst}{t^{\rm 1}}
\newcommand{\tsubs}{t_{\rm s}}
\newcommand{\tsuba}{t_{\rm a}}

\makeatother

\providecommand{\definitionname}{Definition}
\providecommand{\examplename}{Example}
\providecommand{\remarkname}{Remark}
\providecommand{\theoremname}{Theorem}

\begin{document}

\title[Interaction spaces: a universal theory of complex systems]{Interaction spaces: towards a universal mathematical theory of complex
systems}
\author{Paolo Giordano}
\address{Paolo Giordano, Faculty of Mathematics, University of Vienna, AT.}
\thanks{This research was funded in whole or in part by the Austrian Science
Fund (FWF) 10.55776/PAT9221023. For open access purposes, the author
has applied a CC BY public copyright license to any author-accepted
manuscript version arising from this submission}
\begin{abstract}
We present the first steps of \emph{interaction spaces theory}, a
universal mathematical theory of complex systems which is able to
embed cellular automata, agent based models, master equation based
models, stochastic or deterministic, continuous or discrete dynamical
systems, networked dynamical models, artificial neural networks and
genetic algorithms in a single notion. Therefore, interaction spaces
represent a common mathematical language that can be used to describe
several complex systems modeling frameworks. This is the first step
to start a mathematical theory of complex systems. Every notion is
introduced both using an intuitive description by listing lots of
examples, and using a modern mathematical language.
\end{abstract}

\maketitle
\tableofcontents{}

\section{Introduction: why do we need a mathematical theory of complex systems?}

Throughout the history of science, several disciplines have considerably
gained from a sound mathematical foundation: quantum mechanics, continuum
mechanics, thermodynamics, medicine, biology, information science,
economics, social sciences, and urban studies, to name but a few.
Indeed, the contribution of mathematics to many disciplines can be
considered a general process that occurs when the solution of problems
requires the strongest notion of rational truth corroborated by a
meaningful validation.

At present, different modeling methods are adopted to study complex
systems (CS): among the most used, we can cite, e.g., cellular automata
(CA), see e.g.~\cite{Ma-To87}, agent based models (ABM), e.g.~\cite{Woo02},
master equation based models, \cite{Sch02,Haa89}, networked dynamical
systems, \cite{New18}, artificial neural networks, \cite{Hay99,Ma-Ch01,Mi-Ia23},
and genetic algorithms, \cite{Bae96,Mit96}. However, there is no
universal mathematical theory of CS, i.e.~a theory sufficiently powerful
to range over all these systems, from ABM to systems described by
some type of differential equations, and, at the same time, to produce
meaningful general mathematical results applicable to large classes
of systems. The problem is well-known and discussed in literature:
see e.g.~\cite{Hoo11,New11,Ch-St-Fj03,Cru14,Cru03,Bed-etal00,Fon-Bal99,Cas97,Sim97,Hol95a,Far90},
where you can find both opinions in favour or against the possibility
of such a theory.

In this article, we introduce a new mathematical structure, called
\emph{interaction space} (IS), having the property to include (i.e.~to
faithfully embed preserving their original mathematical structure)
in a single notion all the previously listed modeling frameworks.
In our opinion, such a founding mathematical theory could provide
great impact from the perspectives of a common language, precise definitions
and general results which would hence be applicable to all these settings
(see \cite{GucOtt08,Far90,LerGal01} for very similar viewpoints).

Other aims we have in mind are the following:
\begin{itemize}
\item A common mathematical language can be useful to precisely formulate
problems like bifurcations, phase transitions and critical phenomena,
pattern formation theory, ergodic theory, study of ABM as dynamical
systems, etc. (see e.g.~\cite{Cru03,Str14,Fuc13,New11} for similar
problems).
\item With our results on Markovian IS and power law for complex adaptive
IS, we show the possibility to prove general results applicable to
large classes of CS, see \cite{Gio24cas,Gio24nM}.
\item The description of the dynamics of non-Markovian IS with a system
of mean derivative equations, represents a new important general mathematical
result. This demonstrates that using a modern mathematical setting,
powerful mathematical tools can be used to solve open problems, see
\cite{Gio24nM}.
\item IS theory represents a proposal for a sound mathematical definition
of ABM. This definition would open the possibility to start a mathematical
study of a large family of these models (see e.g.~\cite{LerGal01}
for a mathematical approach to the dynamics of some types of ABM).
\item In an IS we also have a language of cause-effect relations, where
elementary modeling-dependent cause-effect relations between interacting
entities can be composed into more complex cause-effect graphs. Using
a suitable language of multicategory theory, these cause-effect relations
can be used both to model hierarchies of complex systems and new general
methods of artificial intelligence, see e.g.~\cite{Pe-Ma18,Gio23}.
See also \cite{BaFo21,FBSD21,BrRoPoSu19} for a similar point of view.
\end{itemize}
In other words, a mathematical theory of CS aims to link phenomenological
studies (e.g.~estimates of power laws) to a modern mathematical theory,
so that to make a step further obtaining more general, clear and widely
applicable results.

\subsection{Other mathematical theories of complex systems}

As far as we know, only the following approaches claim to be mathematical
theories of at least suitable classes of complex systems:
\begin{enumerate}
\item \emph{\label{enu:KTAC}Kinetic theory for active particles}, see \cite{Kno14,AjBeTo13,Cos11,Bel08}
and references therein. This approach is used to describe the dynamics
of a large number of interacting entities in living systems which
are distributed over a network. Usually, entities are homogeneously
distributed within each node and the model provides a mesoscopic description,
i.e.~through the probability distribution over the microscopic states.
The mathematical methods are near to those of statistical mechanics
and game theory.
\item \emph{\label{enu:MES}Memory evolutive systems}, see e.g.~\cite{EhVa87,EhVa14,Kai09}
and references therein. This theory is mainly proposed as a possible
foundation of biology. Deeply based on category theory, it makes extensive
use of limits and colimits of diagrams to model evolving hierarchical
category of living systems. Because of its abstract approach, the
scope of memory evolutive systems is probably very general. In spite
of this abstractness, it captures essential aspects of biological
organization and hence it could lead to concrete hypotheses which
are capable of being tested.
\item \emph{\label{enu:UD}Universal dynamics}, see \cite{Mac01,Mac94,Mac95,Mac96,Mac97,Tom99}.
This approach is also based on category theory, and claims to be a
universal theory for every complex system. The basic structure is
elementary and given by a category with a selected family of arrows,
called fundamental. On the other hand, only local Markovian dynamics
in discrete time is considered because the dynamics depends only on
a finite number of past times. A notion of locality and of neighborhood
is defined using composition of fundamental arrows. In particular,
we underscore its applications in information science in \cite{Tom99,Mac96}.
\item \label{enu:NDD}\emph{Networks and networked dynamical systems}, see
e.g.~\cite{New18,Mit06}. Even if this theory does not usually claim
to be a universal mathematical theory of CS, frequently it is one
of the most effectively used point of view on CS. For example, there
is no general definition of CS nor of complex adaptive system within
this theory, see e.g.~\cite{JaBaCr,EsRo} for arguments supporting
the idea that network theory is insufficient to model several interesting
CS. Moreover, the only network structure of a given model of a CS
does not uniquely determine this model, i.e.~classical models of
CS cannot be identified with their network. As we will see in Sec.~\ref{subsec:Networked-dynamical-systems},
we could say that IS theory can be considered as a more general and
abstract version of this theory, even if it is actually more near
to hypergraphs or multicategories/operads, see e.g.~\cite{EsRo,JaBaCr,BaFo21,FBSD21}.
Indeed, in Sec.~\ref{subsec:Networked-dynamical-systems} we also
prove that every networked dynamical system can be faithfully embedded
as IS.
\end{enumerate}
All these theories, even when they claim to be universal, do not show
clear relations with the most used modeling approaches for CS. For
this reason, one cannot state that their theorems can be applied to
a large family of these models. On the contrary, they present some
limitations, like the mesoscopic, or discrete or Markovian dynamics.
Finally, in our opinion, the abstract approach used both in \ref{enu:MES}
and \ref{enu:UD} sometimes represents an impediment in their spreading
in the scientific community of CS modeling and in their practical
implementation as a computational tool.

In the present work, we see that IS theory includes all classical
models of CS and have a clear cause-effect structure. This allows
us in \cite{Gio24cas} to introduce a meaningful \emph{mathematica}l
notion of complex adaptive system by formalizing informal ideas frequently
used in modeling of CS.

It is important to note that the universality of IS theory allows
one to be sure that sufficiently general mathematical results have
a satisfactorily range of applications for a diverse range of different
modeling frameworks of CS. For theorems already going in this direction,
see \cite{Gio24nM,Gio24cas}. Note that this does not force anyone
to switch to IS from his favorite CS setting, but it only establishes
a general common mathematical language for CS. 

\section{\label{sec:intuitiveInteractionSpaces}Intuitive description of interaction
spaces and their dynamics}

We first describe a generic IS by using only an intuitive approach
and giving several examples, exactly like agent based models (ABM)
are frequently presented. Secondly, we present a mathematical approach,
clearly explaining why this mathematics corresponds to the related
intuitive description.

\medskip{}

IS theory aims at modeling complex systems enclosed in the following
general frame:

\subsection{Interacting entities and their state\label{subsec:Interacting-entities-and-states}}

The system is made by \emph{interacting entities} $e\in E$ described
by dynamical \emph{state variables} $x_{e}(t)$ for $\tst\le t\le\tend\le+\infty$.
Intuitively, an interacting entity is everything able to send or receive
propagator signals (of any type) to interact with other interacting
entities. In general, we think state variables as vectors made of
several components. In case of stochastic dynamics, we can think at
the function of time $x_{e}(-)$ as a sample path followed by the
state of the entity $e$ for some random elementary event $\omega$,
which encloses all the stochastic events from which this dynamics
depends on.\medskip{}

Examples of interacting entities are: agents of an ABM, a vehicle,
a traffic light or the stretch of road between two following cars,
advertisements in a street, goods exchanged in a market, a whole population
of individuals sharing common features and interacting with other
entities, words in a text, cells of a CA (even if in this case the
propagator signals are not considered in the CA model), etc. 

\subsection{Interactions\label{subsec:Interactions}}

\begin{figure}
\noindent \begin{centering}
\includegraphics[scale=0.22]{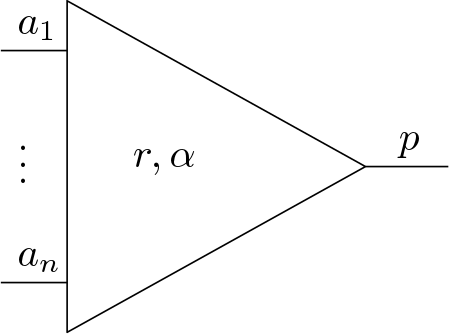}
\par\end{centering}
\caption{\label{fig:diagrInter}Representation of an interaction using a diagram.}
\end{figure}

These interacting entities are involved in interactions $i\in I$,
each one of a given type $\alpha$, that can be described as a causally
directed elementary process in which a set of \emph{agent} entities
$a_{1},\dots,a_{n}$ modify the state of a \emph{patient} entity $p$
through a \emph{propagator} entity $r$. We distinguish between the
type $\alpha$ of the interaction, which is usually a label useful
to classify different interactions, and the interaction $i=(a_{1},\dots,a_{n},r,\alpha,p)$
that includes all these information. The propagator $r$ can be thought
of as a signal-entity activated by agents, and carrying the cause-effect
relation sent by agents $a_{1},\dots,a_{n}$ to the patient $p$.
We also think that a subspace of the state space of the propagator
$r$ works as a \emph{resource space} $R_{i}$ for the changing of
the state of the patient $p$; we will see later why this is important
to define CAS.\\
The general form of an interaction $i$ is hence:
\begin{equation}
i:a_{1},\dots,a_{n}\textsf{ have an interaction }\alpha\textsf{ with }p\textsf{ through }r\label{eq:interactionGeneral}
\end{equation}
which will be also indicated with the notation
\begin{equation}
i:a_{1},\ldots,a_{n}\xra{r,\alpha}p\label{eq:interactionGeneral2}
\end{equation}
or with a diagram as in Fig.~\ref{fig:diagrInter}. This is a sort
of \emph{primitive} \emph{cause-effect relation} (i.e.~it depends
on the constructed model of the considered CS), and our interest lies
more on the possible \emph{cause-effect graphs} that can be built
up by concatenating these elementary relations. In other words, agents
$a_{1},\ldots,a_{n}$ represent the sites of information storage,
and the communication topology of information flows within a system
is explicitly given by propagators in cause-effect interactions such
as \eqref{eq:interactionGeneral2}. See also Sec.~\ref{subsec:Simultaneous-interactions}
for polyadic interactions, as well as \cite{EsRo,JaBaCr,BaFo21,FBSD21}
and references therein for similar viewpoints.

\medskip{}
Examples: a physical interaction between one particle $p_{1}$ sending
a signal $s$ to another particle $p_{2}$, $i=(p_{1},s,\texttt{sendSignal},p_{2})$;
or a firm (agent) sending an advertisement (propagator) and hence
changing the state of several people (patients); a suitable set of
goods in a market (agents) sending a signal (propagator) that carries
information useful for buyers (patients); a biological entity (agents)
sending a chemical signal (propagator) to another entity (patients)
having receptors able to recognize that signal; in a given text, an
adjective $a_{1}$ specifies a name $a_{2}$ hence changing its state
as a patient $p=a_{2}$, and the propagator $r$ can measure the amount
of information specified by the adjective $a_{1}$; an object in an
object oriented program sending a message to another object; a single
neuron has multiple dendrites $a_{1},\ldots,a_{n}$ (inputs from other
neurons), and sends electrical and chemical signals $r$ of type $\alpha$
to its unique axon $p$. In urban models, agents can be individuals
acting in the urban space (e.g.~as builders or residents), patients
can be lots of terrain, propagator signals can be volumes and surfaces
produced for different uses so that the state space of propagators
is linked to the available surface and volume at disposal, depending
on the master plan (which represents the space of resources; see \cite{Va-Gi-An08a,Va-Gi-An08b,AAGV}).
Note that we can have more interactions acting on the same patient,
such as in the case of a car and a pedestrian simultaneously approaching
another pedestrian. We also want to have a sufficient freedom in setting
an IS as a model of a complex system, so that, if needed, we can consider
interacting entities as mathematical idealized entities: for example,
think at a collision between two balls of steel $b_{1}$ and $b_{2}$,
and the possibility to set as propagator the subbody of the Cartesian
product $b_{1}\times b_{2}$ actually involved in the collision. We
can also be interested in considering as ideally infinite the speed
of this propagator in case of elastic collision, so that the aforementioned
subbody is ideally given by the single point of contact.

\subsection{Activation\label{subsec:Activation} }

An interaction $i:a_{1},\ldots,a_{n}\xra{r,\alpha}p$ is occurring
only if \emph{at least one} of the agents $a_{1},\dots,a_{n}$ and
its propagator $r$ is \emph{active for that interaction}. Inded,
in the state $x_{e}(t)$ of each interacting entity $e$ there is
always a time dependent state variable $x_{e}(t)_{1,i}=:\act_{i}^{e}(t)\in[0,1]$
(for simplicity, think at the very common case where $\act_{i}^{e}(t)\in\{0,1\}$
is a Boolean variable) indicating if, with respect to the given interaction
$i$, the entity $e$ is active or not. Intuitively, if the interaction
$i$ starts at time $\tist{i}$, then at least one agent $a_{j}$
must be active with respect to $i$, i.e.~it can be involved in the
interaction $i$, and we have $\act_{i}^{a_{j}}(\tist{i})\ne0$. At
the same starting time $\tist{i}$, agents activate the propagator
$r$: $\act_{i}^{r}(\tist{i})\ne0$. The propagator $r$ will take
a certain time $\tarr{i}-\tist{i}$ to arrive at the patient $p$.
If no other entity and interaction stops $r$ (in that case $\tarr{i}=+\infty$),
$r$ is still active at the arrival time, $\act_{i}^{r}(\tarr{i})\ne0$,
and activates the patient: $\act_{i}^{p}(\tarr{i})\ne0$. See also
Sec.~\ref{subsec:Clock-functions} and Sec.~\ref{subsec:Data-to-run}
for a more accurate formulation of these conditions.\\
Active agents can also be interpreted in biological terms as entities
sending some kind of chemical signal to patients entities having suitable
receptors to recognize it; in this description, propagators are entities
carrying the signal. Therefore, agents which are not already active
in the initial condition of the system, can pass to an active state
as a consequence of an interaction (endogenous or exogenous). Therefore,
both from an intuitive and modelling point of view, the syntactic
structure $i:a_{1},\ldots,a_{n}\xra{r,\alpha}p$ of an interaction
and these activations state variables represent the elementary dynamics
of the cause-effect signals that propagate in the system. These signals
compose themselves into complex cause-effect graphs, whose study is
one of the main interests in modeling CS.

Note that this dynamics of occurrence times and activation functions
represents a stronger formalism with respect to the usual cause-effect
mathematical formalization, as used e.g.~for time series. Indeed,
it is well-known that a simple conditioning can fail to localize information,
so that Shannon entropy and similar measures are not able to measure
information flow, see e.g.~\cite{JaBaCr} and reference therein.

\medskip{}

Examples: in the above mentioned example about firm's advertisement,
only buyers activated, in some way, for the advertised products will
have a state modification; it can also happen that an interaction
of higher priority deactivate a buyer with respect to the advertised
product; only the biological entities having suitable receptors are
active for the corresponding interactions; a computer client is waiting
for a signal from a server before restarting a download, so that it
can be activated at a stochastic future time or deactivated by another
program; a date in a text can activate another specific word, such
as one describing a illness; only software objects with a suitable
public state variable can receive a message to change that variable;
only hungry predators are active for hunting preys, and we can measure
in a fuzzy way $0\le\act_{i}^{e}(t)\le1$ their degree of hungriness;
a similar fuzzy activation can also be useful in suitable models of
Alzheimer disease.

Note that the property of the cause-effect relation \eqref{eq:interactionGeneral2}
of being primitive can also be understood in another way: even if
agents $a_{1},\ldots,a_{n}$ are also intuitively interacting to produce
and activate the propagator $r$, in general we are not interested
to model this kind of more elementary interactions between agents
(think e.g.~at the scattering of two particles or the elastic collision
between two balls where we do not model the dynamics during the collision);
in this case, this means that in our model there are no interactions
of the type $j:a_{k_{1}},\ldots,a_{k_{m}}\xra{s,\beta}a_{h}$ or $l:a_{h_{1}},\ldots,a_{h_{l}}\xra{u,\gamma}r$
prior to $i$. This justifies why $i$ starts at time $\tist{i}$
and, \emph{at the same time}, the propagator is activated: $\act_{i}^{r}(\tist{i})\ne0$.
In other words, interaction between agents $a_{1},\ldots,a_{n}$ happens
in a negligible time span with respect to all the other timings happening
in the system. On the contrary, if we are interested to model the
time used to activate $r$, we have to also consider interactions
of the type $l:a_{h_{1}},\ldots,a_{h_{l}}\xra{u,\gamma}r$ aiming
at activating $r$.

\subsection{Occurrence times\label{subsec:Occurrence-times}}

Once the propagator $r$ arrives at a time $\tarr{i}$, we say that
the interaction $i$ is \emph{ongoing}, in the sense that the state
of the patient $p$ can start to change. If the interaction $i$ is
ongoing at the time $\tong{i}$, this can be instantaneous, i.e.~a
single time instant $\tong{i}=\tarr{i}$, or continuous, i.e.~belonging
to an interval $\tong{i}\in[\tarr{i},\tarr{i}+\delta_{i}]$. Therefore,
whereas the arrival times $\tarr{i}$ of an interaction $i$ are always
single time instants, both the ongoing times $\tong{i}$ and the starting
times $\tist{i}$ can be discrete (e.g.~a discrete dynamical system,
such as a CA) or continuous (e.g.~when agents continuously send the
propagator $r$ to the patient $p$ or its state $x_{p}(t)$ continuously
changes in the interval $[\tarr{i},\tarr{i}+\delta_{i}]$, e.g.~in
a continuous dynamical system). Depending on the considered system
and on its model, all these times $\tist{i}$, $\tarr{i}$ and $\tong{i}$
can be deterministic or stochastic.

More precisely, we can hence think at them as sample paths $\tist{i}=\tist{i}(t)$,
$\tarr{i}=\tarr{i}(t)$ and $\tong{i}=\tong{i}(t)$ for $\tst\le t\le\tend$,
with a suitable model-depending distribution. We always have $\tist{i}(t)\ge t$
(see Sec.~\ref{subsec:Clock-functions}), and $\tist{i}(t)$ can
be thought of as the first starting time of $i$ after or at the present
time $t$. Of course the interaction $i$ can occur multiple times
in $[\tst,\tend]$, and if $\tist{i}(t)=t$ we have that $i$ is starting
exactly at $t$, otherwise that it will start at the time instant
$\tist{i}(t)>t$. See the precise Def.~\ref{def:setOftimeEvents}
and Def.~\ref{def:dataInt}. Similarly, $\tong{i}(t)\ge t$ can be
thought of as the first ongoing instant of time of $i$ after or at
$t$. The arrival time $\tarr{i}(t)$ can actually be defined as the
first of the ongoing times $\tong{i}(t)$ (see Sec.~\ref{subsec:Data-to-run}).
An inequality of the type $\tarr{i}(t)>t$ means that the propagator
$r$ will arrive at a future time $\tarr{i}(t)$; we interpret $\tarr{i}(t)=t$
as the arriving at the present time $t$, and $\tarr{i}(t)<t$ as
the statement that the propagator arrived in the past at $\tarr{i}(t)$.

Therefore, if $\tist{i}(t)=t$, then $\tist{i}(t)\le\tarr{i}(t)$,
i.e.~if the interaction $i$ starts at the present time $t$, then
it will arrive in a future time instant $\tarr{i}(t)\ge\tist{i}(t)=t$
(the propagator cannot arrive in a past time instant $\tarr{i}(t)<\tist{i}(t)$).

The distributions of these times $\tist{i}$, $\tarr{i}$, $\tong{i}$
model the timing of the system, and we can always include the deterministic
cases using suitable Dirac delta distributions, i.e.~using a trivial
probability space.

Clearly, it is because of the universality properties of IS theory
that we aim at this generality. 

\medskip{}
Examples: an interaction where an agent chooses a shop on the basis
of its information about quality, prices, and goods availability,
occurs at random times with a suitable distribution (e.g.~an exponential
distribution whose rate reflects the characteristics of the shop)
depending both on objective and subjective characteristics; an interaction
describing a house leasing occurs at random times depending on several
factors, e.g.\ the rate of birth, of marriage, of immigration, etc;
the infection of an organism by a virus depends randomly on the hosts
encountered; if this virus is considered as the propagator of the
infection interaction, then it will arrive to the possible next organism
after a random time depending on its aging; an excited electron (agent)
produces a photon (propagator) that, in a time depending on the media,
changes the state of another electron (patient) in a scattering interaction;
a word in a text can activate a corresponding mental notion in the
reader; the starting of a program randomly depends on the interaction
of the user with the program's interface. 

\subsection{Neighborhood of an interaction\label{subsec:Neighborhood-of-an}}

The occurrence of an interaction $i$ and its effects depend on the
history of the state of a set of entities $\mathcal{N}_{i}(t)$ called
the \emph{neighborhood} \emph{of the interaction}. The neighborhood
of the interaction $i$ is intuitively defined by all the active entities
from which $i$ takes the information it needs to operate, and it
can depend on time. The neighborhood of an interaction always includes
agent, patient and propagator entities whenever they are active for
that interaction.\medskip{}

Examples: if an agent is searching for a new house, only the information
collected in some order in its memory will affect its future decisions;
only the state of the cells belonging to the neighborhood can influence
the future state of a given cell in a CA; a given negation or an adjective
in a text can influence only a few near verbs or names; only the (random)
objects in the visual field of a pedestrian may influence its goal-oriented
path; the information collected in a graphical user interface may
influence the possible starting of a given computer program.

\subsection{Goods and resources\label{subsec:Goods-and-resources}}

When an interaction $i$ starts at $\tist{i}(t)$, a quantity $\gamma_{i}(t):=x_{r}(t)_{2,i}$
(called \emph{good}) is (probabilistically) extracted from the resource
subspace $R_{i}$ of the propagator $r$. In general, the evolution
of the state variables of the patient $p$ depends on the extracted
goods $\gamma_{i}(t)$. In the space $R_{i}$ we can have a notion
of \emph{zero resources} $Z_{i}\subseteq R_{i}$, so that if $\gamma_{i}(t)\in Z_{i}$,
then $\act_{i}^{p}(t)=0$, i.e.~the patient $p$ is not active for
$i$. This implies that the propagator does not arrive at $t$, i.e.~$\tarr{i}\ne t$
because above we stated that $\act_{i}^{p}(\tarr{i})\ne0$. If, for
a given set of interacting entities (a \emph{population}) these resources
cannot be zero, then other entities in the population will try to
manage this lacking of resources. This is a first very rough explanation
why the notions of goods and resources will be used to define CAS,
see \cite{Gio24cas}.\medskip{}

Examples: an excited electron (agent) produces a photon (propagator)
that changes the state of another electron (patient) in a scattering
interaction, and goods are related to the frequency of the photon.
A specific adjective in a text sends more goods to a given name than
a less specific one. The input currents (propagator) of a neuron are
the signals (goods) that will be integrated to produce a suitable
changing of the output synapses. A developer decides to build a new
house and produces as signal the house's project, hold in the state
of a suitable abstract propagator entity. Starting from this project,
the state of the building's plot will change in a suitable amount
of time, unless the municipal administration blocks the project (resources
are emptied). In general, a situation where the resources are exhausted
before the finishing of the interaction, is an example where the propagator
is deactivated before the ending of the interaction, and hence also
the patient will be deactivated.

\subsection{Evolution equations\label{subsec:Evolution-equations}}

Every model of a CS has corresponding evolution equations satisfied
by the state variables $x_{p}(t)$. These equations can be given by
differential equations, possibly stochastic, or discrete ones; they
can take into account memory effects (i.e.~they are of non-Markov
type) or not, and we need a common language for all of them.\\
Let us consider a patient entity $p\in E$: every model of a CS considers
a transition function $f_{p}$ responsible for the dynamics of the
state $x_{p}(t)$. At the generic present time instant $t\in[\tst,\tend]$,
we consider the first arrival time among all the interactions in our
system that started at $t$ (i.e.~such that $\tist{i}(t)=t$):
\[
\tfirst(t):=\tfirst:=\inf\left\{ \tarr{i}(t)\mid i\in I,\ \tist{i}(t)=t\right\} 
\]
(we read it as \emph{``}$t$\emph{ first}''). Note that, intuitively,
no event occurs in the interval $(t,\tfirst)$. If $\tfirst=+\infty$
this means that no more interactions occur after $t$, so we can assume
$\tfirst<+\infty$. 

Since more than one interaction can simultaneously act on the patient
$p$ during the time interval $[\tfirst,\tfirst+\Delta]$, we consider
all of those interactions, i.e.~all the interactions whose propagator
arrives in this interval
\[
I_{p}(t):=\{i\in I\mid\tfirst(t)\le\tarr{i}(t)\le\tfirst(t)+\Delta,\ \pa{i}=p\}.
\]
Here $\Delta\in\R_{\ge0}\cup\{+\infty\}$ is a model-depending interval
of time representing when the evolution equation defined by $f_{p}$
is solely responsible for the time change of the state $x_{p}(t)$
(see also Rem.~\ref{rem:EE} below for examples and other intuitive
interpretations of $\Delta$).

Note that we have to consider the evolution equation only if $I_{p}(t)$
is not empty because otherwise this would mean that among all the
interactions acting on the patient $p$, no one arrives in the interval
$[\tfirst,\tfirst+\Delta]$.

Now, we can take into account all the non-Markovian dependencies by
considering the \emph{state of the neighbourhood of $p$}:
\begin{align}
 & \text{If }\exists t'\le t\,\exists i\in I:\ \pa{i}=p,\ t'=\tarr{i}(t'),\ \eps\in\mathcal{N}_{i}(t'),\ \tau\in[t',t]\label{eq:neighFunct}\\
 & \text{then }\stateNeigh_{p}x(\tau,\eps):=x_{\eps}(\tau).\nonumber 
\end{align}
Explanation: If in a possible past time $t'\le t$ the propagator
of an interaction $i\in I$ arrived at its patient $p$ (i.e.~$t'=\tarr{i}(t')$),
we consider the state $x_{\eps}(\tau)$ of every interacting entity
$\eps$ in the neighborhood $\mathcal{N}_{i}(t')$ for all the following
times $\tau\in[t',t]$ (see also Rem.~\ref{rem:EE} below for examples
and other intuitive interpretations of this (possibly) non-Markovian
behavior).

The general evolution equation for the patient $p$ can now be stated
as follows: There exists an elementary event $\omega\in\Omega_{p}$
(in a suitable probability space modeling the possible stochastic
evolution of $p$ governed by the evolution equation) such that if
$t\in[\tst,\tend]$, $\tfirst(t)<+\infty$ and $I_{p}(t)$ is not
empty, then for all $s$ such that $\tfirst\le s\le\tfirst+\Delta\le\tend$,
we have
\begin{equation}
x_{p}(s)=f_{p}\left(\omega,s,\stateNeigh_{p}x_{s}\right),\label{eq:EE}
\end{equation}
where $\stateNeigh_{p}x_{s}$ denotes the neighborhood function considered
only in the interval $[\tfirst,s]$, i.e.
\[
\stateNeigh_{p}x_{s}:\tau\in[\tfirst(t),s]\mapsto\stateNeigh_{p}x(\tau,-).
\]
\medskip{}

\begin{rem}
\emph{\label{rem:EE}~}
\begin{enumerate}[label=(\alph*)]
\item In several cases (e.g.~a discrete dynamical system like a CA, where
$\Delta=1$), this $\Delta$ can be thought of as a small interval
of time with respect to the speed at which the changing of the state
$x_{p}(t)$ spread out in the system, and no other interactions occur
in $(\tfirst,\tfirst+\Delta)$.
\item If, in an idealized system, the change of the state $x_{p}(t)$ spread
instantaneously in the whole system, then we have to set $\Delta=0$
and all the interactions occur instantaneously at $\tfirst$.
\item We will see below that in every continuous dynamical system where
$x_{p}(t)$ is described by a differential equation, we can set $\Delta=\tend-\tst$,
because the differential equation governs the evolution of $p$ in
the entire interval $[\tst,\tend]$. Therefore, in this case, $\Delta$
is not small. However, in Sec.~\ref{sec:Classical-models-as-IS},
we will see that for this IS we have only one interacting entity $p$
and only one interaction $i$ corresponding to the differential equation
describing the system.
\item Of course, the dependence on past states expressed by \eqref{eq:neighFunct}
is a very strong one; however, think for example at the case where
$\eps$ represents a malignant tumor diagnosis for the patient $p$
at the time $t'=\tarr{i}(t')$, and all the subsequent ($\tau\in[t',t]$)
medical and psychological consequences on $p$ of the state $x_{\eps}(\tau)$
of the neoplasm $\eps$.
\item \label{enu:EE-ac_goods}Since the state $x_{p}(s)$ includes both
the activation $\act_{j}^{p}(s)=x_{p}(s)_{1,j}$ (see Sec.~\ref{subsec:Activation})
and the goods $\gamma_{j}(s)=x_{p}(s)_{2,j}$ (see Sec.~\ref{subsec:Goods-and-resources}),
the evolution equations \eqref{eq:EE} have to also include their
dynamics:
\begin{align*}
\act_{j}^{p}(s) & =f_{p}\left(\omega,s,\stateNeigh_{p}x_{s}\right)_{1,j}\quad\forall j\in I,\\
\gamma_{j}(s) & =f_{p}\left(\omega,s,\stateNeigh_{p}x_{s}\right)_{2,j}\quad\forall j\in I:\ p=\pr{j}.
\end{align*}
Therefore, these equations also control the cause-effect dynamics
represented by activation states, and the dynamics of goods; the latter
are important for CAS (see \cite{Gio24cas}).
\item \label{enu:EE-dynSys}We will see more precisely later how both continuous
and discrete dynamical systems can be equivalently described using
an equation of the form \eqref{eq:EE}. Here, we only mention that
an ordinary differential equation (ODE) of the form $x'_{p}(s)=F(s,x_{p}(s))$
for all $s\in[\tst,\tend]$ can equivalently be written as $x_{p}(s)=x_{p}(\tst)+\int_{\tst}^{s}F(\tau,x_{p}(\tau))\,\diff\tau=:f_{p}\left(s,x_{p}(-)|_{[\tst,s]}\right)$.
On the other hand, if we have $x_{p}(k+1)=F(k,x_{p}(k))$ for all
$k=0,\ldots,N$ and $x_{p}(0)=x_{0}$, then we can define $f_{p}(s,x_{p}|_{[0,s]})$
stepwise by
\begin{equation}
f_{p}(s,x_{p}|_{[0,s]}):=\begin{cases}
F(k,x_{p}(k)) & \text{if }s=k+1\\
F(k-1,x_{p}(k-1)) & \text{if }s\in[k,k+1)\text{ and }k>0\\
x_{0} & \text{if }s\in[0,1),
\end{cases}\label{eq:DDS}
\end{equation}
to reenter into the language of \eqref{eq:EE}. Note that in both
cases we consider a dependence only on a suitable restriction of $x_{p}$.
\item We could explicitly admit that the intervals $\Delta=\Delta_{p}(t)$
depend both on the patient $p$ and the time $t$. For example, we
can admit that the evolution of $x_{p}(s)$ is described by an ODE
for some $p$ or for certain times $t$, and by a discrete dynamical
system for other $p$ or different times $t$. However, this would
result in more cumbersome notations, and it will never be used in
the present paper.
\item Note that in the function $\stateNeigh_{p}x_{s}$ we have the dependence
from all the interactions $i\in I$ that acted on $p$ in the past.
This dependence is expressed through the states $x_{\eps}(\tau)$
of entities in the neighborhood $\mathcal{N}_{i}(t')$. If the conditions
\eqref{eq:neighFunct} are never satisfied, the evolution function
$f_{p}$ simply does not depend on these states.
\item Only taking the closed interval $[\tfirst,\tfirst+\Delta]$ we can
consider the case $\Delta=0$ and the evolution istantaneosly occurring
at $t=\tfirst$.
\item It is traditional in physics and mathematics to see that the state
variables satisfies some kind of equation attributed to some important
scientist. A minimal thinking allows us to say that it is ingenuous
to believe that this could happen for all possible complex systems.
We are focusing more on a universal mathematical language. CAS and
the GEP could play the role of this general law, see \cite{Gio24cas},
but not in the simple form of an equation.
\end{enumerate}
\end{rem}

\medskip{}

The following examples surely can be described in the previous intuitive
formalism: a bouncing billiard ball; a pedestrian between two subsequent
interactions with other pedestrians or obstacles; the process of building
a house after its starting time and before its end; the internal evolution
of a box in a flow chart representing a computer program; the patient
$p$ represents a company listed on the stock exchange, and $\eps$
represents another company selling the same type of product which
experienced a strong decreasing of its shares at time $t'$. The interacting
entity $\eps$ represents a neoplasm appearing in a person $p$ at
time $t'\le t$ but still interacting with $p$ at present time $t$.
The last system is clearly non-Markovian, as one can see comparing
two different samples paths where $x_{\eps}(t')=\text{benign}$ or
$\bar{x}_{\eps}(t')=\text{malignant}$.\medskip{}

\begin{figure}
\noindent \begin{centering}
\includegraphics[scale=0.75]{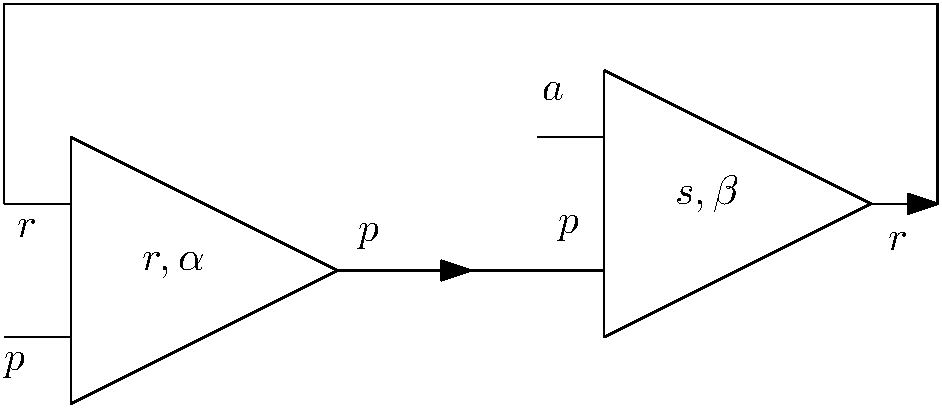}
\par\end{centering}
\caption{\label{fig:interaction1}Graphical representation of two interactions
$i:r,p\xra{r,\alpha}p$ and $j:a,p\xra{s,\beta}r$, where the interacting
entity $r$ is at the same time agent and propagator of the first
interaction $i$ and patient of the second interaction $j$. Using
the agent $a$, we can change the status of $r$ and hence the goods
$\gamma_{i}(t)=x_{r}(t)_{2,i}$ of the first interaction $i$.}
\end{figure}

The terms agent, patient, propagator and members of a neighborhood
are collectively named \emph{roles} of entities in an interaction.
Of course, interacting entities can play different roles in different
interactions and more than one role in the same interaction, e.g.~a
propagator of $i$ can also be at the same time an agent of the same
interaction and a patient of another interaction $j$ which triggers
the goods of $i$. Therefore, if we represent an interaction by means
of a graph, like in figure \ref{fig:interaction1}, and connect two
graphs when they share an entity, we obtain a network representing
the mentioned causal flows in the system. Note that this informal
description already allows for a practical implementation of simulated
IS (see e.g.~\cite{Va-Gi-An14}).

The intuitive description above can be summarized by saying: in an
interaction, agents activate and the propagator and the goods are
sent as a signal to modify the state of the patient; the modification
depends on information collected from the neighborhood of that interaction;
the starting time and the speed of the signal of the interaction can
be stochastic. Occurrence of interactions is causally constrained
by logical conditions expressed by the activation of the entities.
All the interactions acting on patients cause the evolution of their
state during sufficiently small intervals with respect to the spreading
of these changes in the system.

Finally note that if we aim to describe ``a general CS'', terms
such as \emph{interacting entities}, \emph{interactions}, being \emph{active}
or not for an interaction, \emph{neighborhood} as the set of all the
entities where an interaction takes all the needed information, \emph{occurrence
times} and \emph{evolution equations} seems very natural and necessary
notions.

\subsection{\label{subsec:Dynamics-of-IS}Dynamics of an interaction space}

Similarly to an asynchronous CA, the dynamics of a generic IS, is
determined by the occurrence times $\tist{i}$, $\tarr{i}$ and $\tong{i}$
(see Sec.~\ref{subsec:Occurrence-times}) of all the interactions
$i$, and by the evolution equations \eqref{eq:EE}, starting from
an initial state of the system:
\begin{enumerate}[leftmargin=*,label=(\alph*),align=left ]
\item \label{enu:initialState}The system starts with a given initial value
of all the states $x_{e}(\tst)$ for each interacting entity $e\in E$.
Note that this includes the initial values of the activation states
$\act_{i}^{p}(\tst)=x_{p}(\tst)_{1,i}$ and of the goods $\gamma_{i}(\tst)=x_{p}(\tst)_{2,i}$
(if $p=\pr{i}$) for any interaction $i\in I$.
\item For each interaction $i\in I$, we have to provide the starting time
$\tist{i}(\tst)\ge\tst$. If the interaction $i$ starts at $\bar{t}:=\tist{i}(\tst)$,
we also have to provide the arrival time $\tarr{i}(\bar{t})\ge\tist{i}(\bar{t})$
and the ongoing time $\tong{i}(\bar{t})\ge\tarr{i}(\bar{t})$. These
are modeling-depending quantities, and frequently they are random
variables depending on the state of the neighborhood of $i$ at $t=\tst$,
i.e.~on the function $\stateNeigh_{i}x(\tst,\eps):=x_{\eps}(\tst)$
for all $\eps\in\mathcal{N}_{i}(\tst)$. For an arbitrary time $t\in[\tst,\tend]$,
these occurrence times can also depend on past time states $\stateNeigh_{i}x(t',\eps):=x_{\eps}(t')$
for all $\eps\in\mathcal{N}_{i}(t')$ and all $t'\in[\tst,t]$.
\item We compute the first arrival time $\tfirst$. Assuming, for simplicity,
that we have a finite number of interactions, this is given by
\[
\tfirst(\tst):=\tfirst:=\min\left\{ \tarr{i}(\tst)\mid i\in I,\ \tist{i}(t)=t\right\} .
\]
If $\tfirst=\tst$, this means that at least one propagator instantaneously
arrives at $\tist{i}(\tst)=\tarr{i}(\tst)=\tst$. Otherwise, $\tfirst>\tst$
and hence all the states remain constant at $x_{e}(\tst)$ for all
$t\in[\tst,\tfirst)$ because only the evolution equations \eqref{eq:EE}
can change these states. If $\tfirst=t_{i}^{a}(\tst)$ for some $i\in I$,
the occurrence times must coherently satisfy $\bar{t}=\tist{i}(\bar{t})\le\tfirst=t_{i}^{a}(\bar{t})$
for some $\bar{t}\ge\tst$ when $i$ actually started ($\bar{t}=\tist{i}(\bar{t})$),
$\act_{i}^{a_{j}}(\bar{t})\ne0\ne\act_{i}^{r}(\bar{t})$ for some
agent $a_{j}$ and for the propagator $r$ of $i$, and $\act_{i}^{r}(\tfirst)\ne0\ne\act_{i}^{p}(\tfirst)$
for the patient $p$ (see Sec.~\ref{subsec:Activation}).
\item If $\tfirst<+\infty$, then this $\tfirst$ is the first time corresponding
to the arrival of some propagator, and we can hence update the time
as $t=\tfirst$, in the sense that nothing occurred before. Using
the evolution equations \eqref{eq:EE}, we can change the state $x_{p}(s)$
of all the patients of interactions $i\in I_{p}(\tfirst)$ whose arrival
time $\tarr{i}\in[\tfirst,\tfirst+\Delta]$.
\item \label{enu:aPatientIsChanged}Because we have changed the state of
these patient entities, we recursively restart from the beginning
with the new states $x_{e}(t)$ at $t=\tfirst+\Delta$.
\item \label{enu:constantParameters}Note that if an interacting entity
is never a patient of at least an interaction $i$ whose propagator
actually arrives at some $t=\tarr{i}(t)\le\tend$, then its state
$x_{e}(t)$ will never change, because we can never apply the evolution
equation \eqref{eq:EE}. In this case, the states $x_{e}$ of these
kind of interacting entities work as constant parameters of the system.
\end{enumerate}
We can therefore say that \emph{mathematically solving an IS} means
setting the model by deciding interacting entities and interactions,
occurrence times, neighborhoods of interactions, initial states $x_{e}(\tst)$
and transition functions $f_{p}$, and solve or simulate the evolution
equations for the states $x_{p}(-)$. For some models, the occurrence
times or the neighbourhoods can also be considered as unknowns of
the study.

More generally, \emph{solving an IS} means:
\begin{enumerate}[leftmargin=*,label=\arabic*),align=left ]
\item Mathematically solving it;
\item Validating the obtained results by comparison with independent real
world data.
\item This comparison is based on a notion of truth which is accepted by
a certain community at a certain time.
\item This notion have to include understanding and showing frameworks where
the model can and where it cannot be applied (falsification).
\item Careful checking that the model is applied only in validated settings
satisfying all the modeling assumptions.
\end{enumerate}

\section{\label{sec:MathDefOfIS}Mathematical definition of IS}

A mathematical definition of IS is a necessary step to start a mathematical
theory, and hence to prove general theorems in a clear way and using
modern and advanced mathematical instruments. We already started this
process, e.g.~by showing a general master equation for Markov IS
(see \cite{Gio24nM}), proving a systems of mean derivative equations
for the description of a general class of non-Markov IS, \cite{Gio24nM},
giving a very comprehensive definition of CAS (see \cite{Gio24cas})
and proving related power laws, \cite{Gio24cas}.

The usefulness of this mathematical formalization can also be inferred
by thinking at the same basic notions of IS, with cause-effect elementary
relations represented by interactions \eqref{eq:interactionGeneral2}
and activation states. Indeed, these concepts allow one to define
the notion of cause-effect graphs occurring in a system, and of hierarchical
functors that preserve such relations between pairs of different IS.
Therefore, this direction of theoretical development, which we postpone
to a subsequent article, finds potential applications in several CS
such as the brain and more general intelligent systems, urban systems,
the immune system, organisms in biology, social systems, etc. and
wherever the intelligibility of a system using cause-effect graphs
or a hierarchical description can be helpful, see e.g.~\cite{Pe-Ma18,Gio23}.

\subsection{\label{subsec:Interacting-entities-and-interactions}Interacting
entities and interactions }

Already in the informal description of IS, it is clear that many components
are needed to define an IS: a set of entities, a set of types of interactions,
state maps, occurrence times, etc. For this reason, using a nested
approach, we introduce four structures that will define the notion
of IS. In this way, instead of referring only to the complete notion
of IS, we can also focus on only some of these structures and thereby
considering more general modeling settings. 
\begin{defn}
\label{def:EntitiesAndInteractions}A \emph{system of entities and
interactions} $\mathcal{EI}=(E,\tst,\tend,\mathcal{T},I)$ is given
by the following data which satisfy the following conditions:
\begin{enumerate}
\item A set $E$, called \emph{the set of interacting entities}.

\item A\emph{ time interval} $[\tst,\tend]$, with $\tst<\tend\le+\infty$.
\item A finite set $\mathcal{T}$ called \emph{the set of types of interactions}.
\item \label{enu:arietyOfInteractions}A set $I$ called \emph{the set of
interactions} satisfying the following condition: every interaction
$i\in I$ can be written as $i=(a_{1},\dots,a_{n},r,\alpha,p)$ for
some type of interaction $\alpha\in\mathcal{T}$, some entities $a_{1},\dotsc a_{n}$,
$r$, $p\in E$, and where also $n\ge0$ depends on $i$.
\end{enumerate}
\end{defn}

\medskip{}

\begin{rem}
~
\begin{enumerate}[label=(\alph*)]
\item We set $E_{i}:=\left\{ a_{1},\dots,a_{n},r,p\right\} $, $\ag{i}:=(a_{1},\ldots,a_{n})$,
$\pa{i}:=p$ and $\pr{i}:=r$ to denote all the interacting entities
involved in the interaction $i$, agents, patient and propagator of
$i$, resp. For example, if $i=(a,b,b,\alpha,b)\in I$, where $a$,
$b\in E$ and $\alpha\in\mathcal{T}$, this means (reading backwards)
that $\pa{i}=b$, the interaction $i$ is of type $\alpha$, $\pr{i}=b$,
and $\ag{i}=(a,b)$, so that $n=2$.
\item What are naturally thought of as agents or patient can depend on a
fixed frame of reference: think e.g.~at the interaction of collision
between two particles and a frame at rest with respect to one of the
two, which can be naturally thought of as the patient of the collision.
\item There is no a priori limitation on the cardinality of the set $E$
of interacting entities, even though in several cases it is finite.
\item The system is studied in the time interval $[\tst,\tend]$; clearly,
if $\tend=+\infty$, we will use the notation $[\tst,\tend]=[\tst,+\infty]$
to mean $[\tst,+\infty)$.
\item Generally speaking, the interactions are non Newtonian: they involve
more than one agent and they are, in general, not reversible, i.e.~there
is not an action-reaction principle. For example, it does not seem
useful to think as Newtonian the non-colliding interaction of a pedestrian
with an obstacle or the interaction of a builder with a house under
construction or of an object in an object oriented programming language
with another object: even if frequently to each one of these interactions
correspond another interaction as answer, in general there is no useful
way to say that the intensity (force) of the cause interaction is
the opposite of the intensity (force) of the reaction interaction.
\end{enumerate}
\end{rem}

\subsection{\label{subsec:State-spaces-and-activations}State spaces and activations}

As we already intuitively explained in Sec.~\ref{subsec:Interacting-entities-and-states},
Sec.~\ref{subsec:Activation} and Sec.~\ref{subsec:Goods-and-resources},
each interacting entity is described by a state variable $x_{e}$,
of which activation $\act_{i}^{e}$ and goods $\gamma_{i}$ are particular
cases. Goods $\gamma_{i}$ are taken from a subspace $\gamma_{i}\in R_{i}$
of the state space called space of resources of an interaction. The
main aim of the next definition is to specify, from the mathematical
point of view, the entire state space of an interacting entity $e$,
and to underscore that both activation and goods are state variables.

In the following, if $X$ and $Y$ are two sets, $Y^{X}$ denotes
the space of all the functions $f:X\ra Y$ and for the values $f(x)\in Y$
we can also sometime use the notation $f_{x}\in Y$. For the sake
of clarity: if the index set $J=\{j_{1},\ldots,j_{n}\}$ is finite,
then the product of sets is $\prod_{j\in J}S_{j}=S_{j_{1}}\times\ldots\times S_{j_{n}}$.
\begin{defn}
\label{def:stateSpacesActivation}Let $\mathcal{EI}=(E,\tst,\tend,\mathcal{T},I)$
be a system of entities and interactions. A \emph{system of states
}$\mathcal{S}=(S,\mathfrak{S},R,x)$\emph{ }for\emph{ $\mathcal{EI}$}
is given by the following data which satisfy the following conditions:
\begin{enumerate}
\item For every interacting entity $e\in E$, a measurable space $(S_{e},\mathfrak{S}_{e})$
called the\emph{ proper state space }of the interacting entity $e$.
\end{enumerate}
\begin{enumerate}[resume, leftmargin=*,label=(ST),align=left ]
\item \label{enu:ST}For each interacting entity $e\in E$ and time $t\in[\tst,\tend]$,
a \emph{state function} 
\[
x_{e}(t)\in[0,1]^{I}\times\prod_{\substack{i\in I\\
e=\pr{i}
}
}\!\!\!R_{i}\,\,\times S_{e}=:\bar{S}_{e}
\]
\end{enumerate}
\begin{itemize}
\item This means that $x_{e}(t)$ has three components: the first one $x_{e}(t)_{1}\in[0,1]^{I}$
is a function $x_{e}(t)_{1}:I\ra[0,1]$ and its evaluation at $i\in I$
is denoted by $\act_{i}^{e}(t):=x_{e}(t)_{1,i}\in[0,1]$ and called
\emph{activation} of $e$ for the interaction $i$ at time $t$.
\item The second component $x_{e}(t)_{2}$ is defined only if $e=\pr{i}$
is a propagator of some interaction $i$ (otherwise, it is not defined).
Therefore, in general, if $i:a_{1},\ldots,a_{n}\xra{r,\alpha}p$ is
an interaction, we set $\gamma_{i}(t):=x_{r}(t)_{2,i}\in R_{i}$,
and call $R_{i}$ the \emph{space of resources} of $i$. This state
variable $\gamma_{i}(t)$ is called \emph{goods} of $i$.
\item The third component $x_{e}(t)_{3}\in S_{e}$ lies in the proper state
space $S_{e}$. Since we use the specific notations $\act_{i}^{e}(t)$
and $\gamma_{i}(t)$ for the first two components, it is not confusing
using simply the classical notation $x_{e}(t)\in S_{e}$ for the third
one.
\end{itemize}
\end{defn}

\medskip{}

\begin{rem}
~
\begin{enumerate}[label=(\alph*)]
\item The property of the proper state space $(S_{e},\mathfrak{S}_{e})$
of being a measurable space is very weak, from the mathematical point
of view, even if usually on a state space there is a richer structure,
e.g.~$S_{e}=\R^{d}$ for some $d>0$ depending on $e\in E$. Let
us note explicitly that the state space is not time dependent.
\item We say that the interaction $i:a_{1},\ldots,a_{n}\xra{r,\alpha}p$
\emph{has a notion of zero resources} $Z_{i}$ (see Sec.~\ref{subsec:Goods-and-resources})
if
\begin{enumerate}
\item $\emptyset\ne Z_{i}\subseteq R_{i}$;
\item $\forall t\in[\tst,\tend]:\ \gamma_{i}(t)\in Z_{i}\ \Rightarrow\ \act_{i}^{p}(t)=0$,
i.e.~whenever $\gamma_{i}(t)\in Z_{i}$, the patient $p$ is not
active for $i$, i.e.~$\act_{i}^{p}(t)=0$.
\end{enumerate}
\end{enumerate}
\end{rem}

The label \ref{enu:ST} recalls \emph{state function}.

\subsection{\label{subsec:Clock-functions}Clock functions}

In the present section, we want to clarify that both the starting
time $t\in[\tst,\tend]\mapsto\tist{i}(t)$ and ongoing time functions
$t\in[\tst,\tend]\mapsto\tong{i}(t)$ satisfy similar general properties.

Thinking at the dynamics of an IS presented in Sec.~\ref{subsec:Dynamics-of-IS},
we can understand that this dynamics is event based, driven by starting
and arrival times of propagators of interactions. This is the concrete
notion of time as it naturally plays in an IS. Stating it differently:
if we have an interaction sending a propagator at every ticking of
a clock, then ``Time is what clock shows'', as Einstein is supposed
to have said. Conceptually, this is different from the quantity $t\in[\tst,\tend]$,
which is only an independent variable to mathematically manage functions
such as $t\in[\tst,\tend]\mapsto\tist{i}(t)\in[\tst,\tend]\cup\{+\infty\}$.

We start by defining what are the set of time events $T$ we consider
in every IS. We can think $T$ as the stochastic values of an exponential
distribution representing the intensity of occurrence of a given interaction,
see Sec.~\ref{subsec:Stochastic-generationClocks}, or a time interval
$[t^{1},t^{2}]\subseteq[\tst,\tend]$ used to model a continuous dynamical
system. Each one of these $T$ defines a clock function:
\begin{defn}
\label{def:setOftimeEvents}We say that $T$ is \emph{a set of discrete
or continuous time events} (we briefly write \emph{$T$ discr./cont.})
if:
\begin{enumerate}
\item \label{enu:discr_cont}$T\subseteq[\tst,\tend]$ is the disjoint union
of single instants $t_{j}$ for $j\in N\subseteq\N$, or of intervals
$[t_{k}^{1},t_{k}^{2}]$ for $k\in M\subseteq\N$.
\item \label{enu:Accumulation-points}Accumulation points of $T$ lie only
in its subintervals, i.e.
\[
\forall t\in T'\,\exists k\in M:\ t\in[t_{k}^{1},t_{k}^{2}].
\]
We recall that the set of accumulation points is defined by $t\in T'$
if $\forall\delta\in\R_{>0}\,\exists\bar{t}\in T\cap(t-\delta,t+\delta):\ \bar{t}\ne t$,
i.e.~it is the set $T'$ of all the points that can be arbitrarily
approximated using points of $T$.
\end{enumerate}
Moreover, we say that $\tau$ is a \emph{clock function} if
\begin{enumerate}[resume]
\item \label{enu:funClock}$\tau:[\tst,\tend]\ra[\tst,\tend]\cup\{+\infty\}$.
\item \label{enu:infClock}$\exists T\text{ discr./cont.}\,\forall t\in[\tst,\tend]:\ \tau(t)=\inf\left\{ s\ge t\mid s\in T\right\} $.
\end{enumerate}
\end{defn}

\noindent We explain the motivations of this definition in the following

~
\begin{rem}
~
\begin{enumerate}[label=(\alph*)]
\item Written explicitly, condition \ref{enu:discr_cont} is $T=\bigcup_{j\in N}\{t_{j}\}\cup\bigcup_{k\in M}[t_{k}^{1},t_{k}^{2}]$,
and all these unions are disjoint.
\item Condition \ref{enu:Accumulation-points} excludes situations such
as $T=\left\{ 1\pm\frac{1}{n}\mid n\in\N_{>0}\right\} \cup\{1\}$
where it is not clear whether at $t=1$ an event occurs instantly
or continuously.
\item Condition \ref{enu:infClock} can be interpreted saying that $\tau(t)$
is the next time event in $T$ after or at $t$. On the other hand,
when we have $\tau(t)=t$, we say that $\tau$ \emph{is occurring
at} $t$: e.g.~if $\tist{i}(t)=t$, we say that the interaction $i$
is starting at $t$; if $\tist{i}(t)>t$, we say that after $t$ the
interaction $i$ will start the first time at the time instant $\tist{i}(t)$.
\item In figure \ref{fig:ClockEx}, we represented in red the clock function
corresponding to the discr./cont. time events $T$ depicted in blue
on the $y$-axis. In this $T$, at $\tst$, $t_{1}$ and $\tend$
instantaneous events occur, whereas we have a continuous one in $[t_{1}^{1},t_{1}^{2}]$.
\end{enumerate}
\end{rem}

We have the following general results, which can be easily proved
from the previous definitions.
\begin{thm}
\label{thm:clock}If $\tau$ is the clock function defined by the
discr./cont. set of events $T$, then:
\begin{enumerate}
\item \label{enu:minMax}$\exists\min(T)$, $\max(T)$ and $\tau(\tst)=\min(T)\in T$.
\item If $\max(T)=\tend$, then $\tau(\tend)=\tend$, otherwise $\tau(t)=+\infty$
for all $t\in(\max(T),\tend]$.
\item For all $t$, we have $\tau(t)\ge t$, and $\tau(t)=t$ if $t\in T$.
\item The function $\tau$ is non-decreasing, and hence $\tau(t)=\inf\left\{ \tau(s)\mid s\in[t,\tend]\right\} $
for all $t$.
\item If $t_{2}\in T$, $t_{1}<t_{2}$ and $(t_{1},t_{2})\cap T=\emptyset$,
then $\tau(t)=t_{2}$ for all $t\in[t_{1},t_{2}]$, i.e.~$\tau$
is constant and left-continuous in this interval.
\item If $\max(T)=\tend$, then $T=\tau\left([\tst,\tend]\right)$.
\item If $\max(T)<\tend$, then $T\cup\{+\infty\}=\tau\left([\tst,\tend]\right)$.
Therefore, this and the previous property show that the function $\tau$
uniquely determines the set of events $T$ as $T=\tau\left([\tst,\tend]\right)\setminus\{+\infty\}$,
so that we can equivalently work with $T$ or $\tau$.
\end{enumerate}
\end{thm}

\begin{center}
\begin{figure}
\begin{centering}
\includegraphics[scale=0.2]{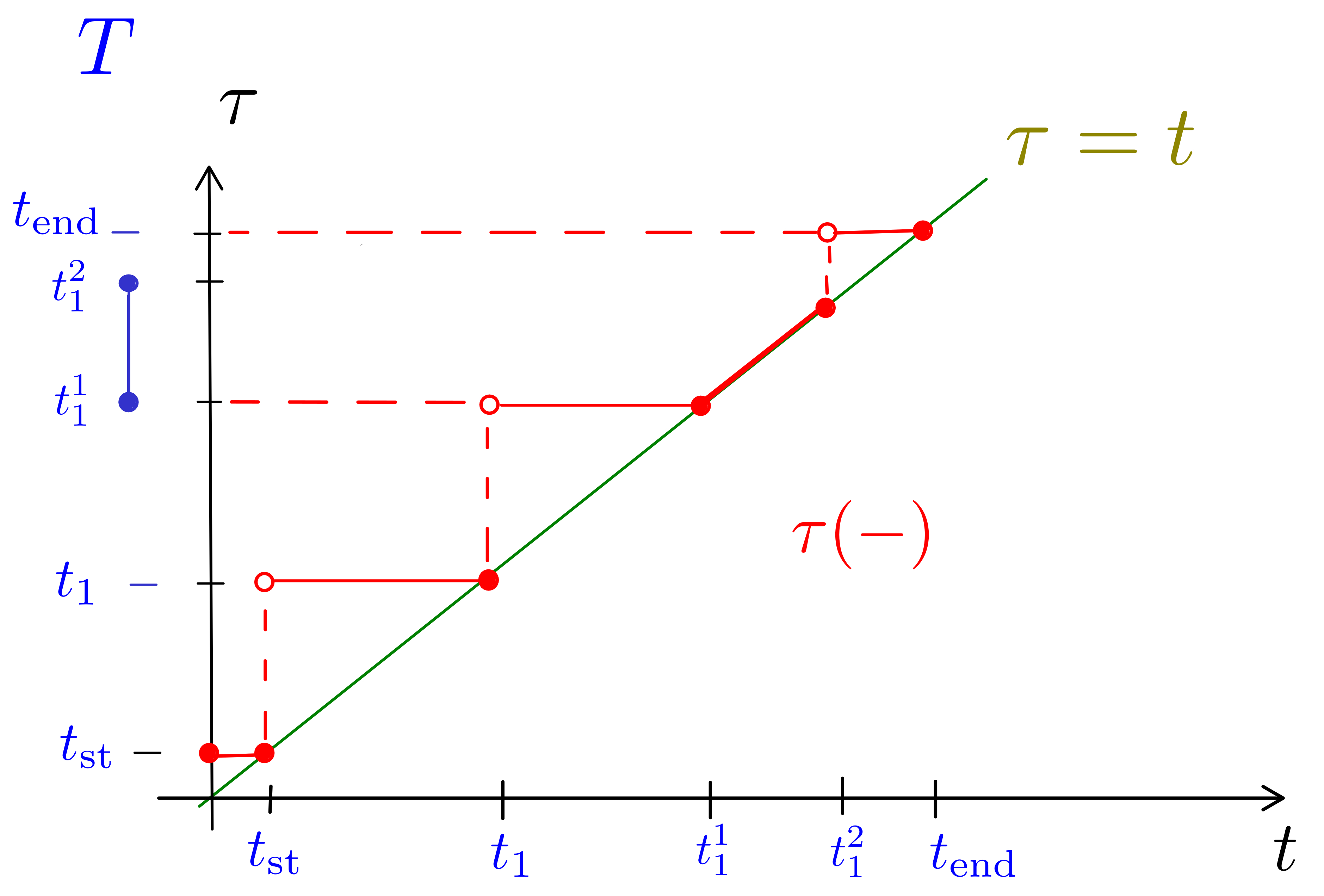}
\par\end{centering}
\caption{\label{fig:ClockEx}An example of clock function.}

\end{figure}
\par\end{center}

\subsection{\label{subsec:Data-to-run}Data to run an interaction}

The previous section gives us the language to formulate the intuitive
statements we already introduced in Sec.~\ref{subsec:Activation},
\ref{subsec:Occurrence-times}, \ref{subsec:Neighborhood-of-an}.
We also want to show that the arrival time function $t\in[\tst,\tend]\mapsto\tarr{i}(t)$
can be defined as the minimal value of $\tong{i}(t)$.
\begin{defn}
\label{def:dataInt}Let $\mathcal{EI}=(E,\tst,\tend,\mathcal{T},I)$
be a system of entities and interactions, and let $\mathcal{S}=(S,\mathfrak{S},R,x)$
be a system of states for \emph{$\mathcal{EI}$. Let} $i:a_{1},\ldots,a_{n}\xra{r,\alpha}p$
be an interaction in $I$, then $\mathcal{D}_{i}=(\tist{i},\tong{i},\mathcal{N}_{i})$
are \emph{data to run} $i$ if:
\begin{enumerate}[resume, leftmargin=*,label=(CF),align=left ]
\item \label{enu:CF}$\tist{i}$ and $\tong{i}$ are clock functions, called
resp.~\emph{starting times} and \emph{ongoing times} of $i$.
\end{enumerate}
Recalling that the set of values $T^{\text{o}}:=\tong{i}\left([\tst,\tend]\right)\setminus\{+\infty\}$
is the set of discr./cont. time events of the ongoing function $\tong{i}$,
we define the \emph{arrival times} of $i$ as follows:
\begin{itemize}
\item $\tarr{i}(t):=t_{j}$, if $\tong{i}(t)=t_{j}$ is discrete (i.e.~$i$
occurs instantaneously at $t=t_{j}$).
\item $\tarr{i}(t):=t_{k}^{1}$, if $\tong{i}(t)\in[t_{k}^{1},t_{k}^{2}]\subseteq T^{\text{o}}$
(i.e.~$i$ occurs continuously around $t$).
\item $\tarr{i}(t):=+\infty$, if $\tong{i}(t)=+\infty$ (i.e.~$i$ never
occurs at $t$ or after).
\end{itemize}
\begin{enumerate}[resume, leftmargin=*,label=(SA),align=left ]
\item \label{enu:SA}For all $t\in[\tst,\tend]$, if $\tist{i}(t)=t$,
i.e.~the interaction $i\in I$ is starting at $t$, then $\tist{t}(t)\le\tarr{i}(t)$.
In other words, if $i$ starts at $\tist{t}(t)=t$, then the propagator
cannot arrive before this starting time, i.e.~$\tist{t}(t)>\tarr{i}(t)$
cannot happen. 
\end{enumerate}
We say that $(t_{\text{s}},t_{\text{a}})$ are \emph{start-arrival
events for $i$ at $t\in[\tst,\tend]$}, if $t_{\text{s}}=\tist{i}(t)=t\le t_{\text{a}}=\tarr{i}(t)<+\infty$.
These data have to satisfy the following conditions:
\begin{enumerate}[resume, leftmargin=*,label=(CE),align=left ]
\item \label{enu:CE}$\act_{i}^{a_{j}}(\tsubs)\ne0$ for some $j=1,\ldots,n$,
$\act_{i}^{r}(\tsubs)\ne0$, $\act_{i}^{r}(\tsuba)\ne0$ and $\act_{i}^{p}(\tsuba)\ne0$.
\end{enumerate}
Finally, the \emph{neighborhood }function $\mathcal{N}_{i}$ satisfies
\begin{enumerate}[leftmargin=*,label=(NE),align=left ]
\item \label{enu:NE}$\left\{ e\in E_{i}\mid\act_{i}^{e}(t)\ne0\right\} =:E_{i}^{t}\subseteq\mathcal{N}_{i}(t)\subseteq E_{t}:=\left\{ e\in E_{i}\mid\exists i\in I:\ \act_{i}^{e}(t)\ne0\right\} $
for all $t\in[\tst,\tend]$, i.e.~the neighborhood $\mathcal{N}_{i}(t)$
always contains the entities of $i$ which are active at $t$, and
it also always contains only active entities.
\end{enumerate}
\end{defn}

The labels \ref{enu:CF}, \ref{enu:SA}, \ref{enu:CE} and \ref{enu:NE}
recall \emph{clock functions}, \emph{start-arrival}, \emph{cause-effect}
and \emph{neighborhood} respectively.

Let us assume, e.g., that $\tist{i}(0)=0=\tst$ and $\tong{i}(t)=1$
for $t\in[0,1]$ and $\tong{i}(t)=t$ for $t\in[1,2]$. We have $\tarr{i}(t)=1$
for all $t\in[0,2]$. We can also have $\tist{i}(t)=3$ for $t>0$,
i.e.~after $0$ the interaction $i$ will start again at $t=3$.
Note that in this case $\tong{i}(1)=1<t_{i}^{s}(1)=3$ which simply
means that $i$ is ongoing at $t=1$ and it will start again at $t=3$.
Therefore, in general, the inequality $\tist{i}(t)\le\tong{i}(t)$
does not hold. Moreover, $t_{i}^{a}(2)=1<2$, therefore the arrival
function does not satisfy $\tarr{i}(t)\ge t$ and hence it is not
a clock function. On the other hand, note that at $t=0$, we have
$\tist{i}(0)=0$, i.e.~$i$ starts at $t=0$, and $\tarr{i}(0)=1>\tist{i}(0)$.
Therefore, also the inequality $\tist{i}(t)\le\tarr{i}(t)$ in general
does not hold (compare this with \ref{enu:SA}).

Since we think at the activation state variable $\act_{i}^{e}(t)=x_{e}(t)_{1,i}$
as a (possible) stochastic path of our CS, from \ref{enu:NE} it follows
that also $\mathcal{N}_{i}(t)\subseteq E_{t}$ has to be thought of
as a (possible) stochastic set.

From the previous definition, we have the following
\begin{thm}
\label{thm:startArrNeigh}In the previous assumptions, if $t\in[\tst,\tend]$,
we have:
\begin{enumerate}
\item \label{enu:tarrPiecewiseConst}$\tarr{i}:[\tst,\tend]\ra[\tst,\tend]\cup\{+\infty\}$
is piecewise constant.
\item $\tarr{i}(t)\le\tong{i}(t)$.
\item If $i:a_{1},\ldots,a_{n}\xra{r,\alpha}p$ is an interaction in $I$,
and $(t_{\textrm{\emph{s}}},t_{\text{\emph{a}}})\ge t$ are start-arrival
events for $i$, then:
\begin{enumerate}
\item $\exists j=1,\ldots,n:\ a_{j}\in\mathcal{N}_{i}(\tsubs)$.
\item $r\in\mathcal{N}_{i}(\tsubs)$ and $r$, $p\in\mathcal{N}_{i}(\tsuba)$.
\item If $i$ has a notion $Z_{i}$ of zero resources, then $\gamma_{i}(\tsuba)\notin Z_{i}$.
\end{enumerate}
\item For all $i\in I$ and $t\in[\tst,\tend]$, there exists $t'\in[\tst,t]$
such that $\tist{i}(t')=t'\le\tarr{i}(t')$, i.e.~there exists a
time $t'\le t$ when $i$ started.
\end{enumerate}
\end{thm}

\subsubsection{\label{subsec:Simultaneous-interactions}Simultaneous vector interactions}

If two interactions $i:a_{1},\ldots,a_{n}\xra{r,\alpha}p$ and $j:b_{1},\ldots,b_{m}\xra{s,\beta}q$
act on patients $p$, $q$ and are \emph{simultaneous}, i.e.~they
have the same occurrence times clock functions $\tist{i}(-)=\tist{j}(-)$
and $\tong{i}(-)=\tong{j}(-)$, we can define a \emph{vector interaction
$(i,j)$} by simply considering as agents
\[
\ag{i,j}:=(a_{1},\ldots,a_{n},b_{1},\ldots,b_{m}),
\]
as patient $\pa{i,j}:=(p,q)$, as propagator $\pr{i,j}:=(r,s)$ and
as type $(\alpha,\beta)$. The activation maps of agents, propagator
and patient are defined in a natural way as $\act_{(i,j)}^{e}(t):=\act_{i}^{e}(t)\cdot\act_{j}^{e}(t)$.
The resource space of $(i,j)$ is the product of the resources of
its components $R_{(i,j)}=R_{i}\times R_{j}$. Similarly, we can define
the state space of the new patient entity $\pa{i,j}=(p,q)$ and the
neighborhood. In the particularly interesting case when the two interactions
act on the same patient $p=q$, we simply set $\pa{i,j}=p$. At the
end, we obtain a cause-effect simultaneous interaction of the form
\[
(i,j):a_{1},\ldots,a_{n},b_{1},\ldots,b_{m}\xra{(r,s),(\alpha,\beta)}p,q.
\]

It clearly depends on our modeling aims whether the  interaction $(i,j)$
already lies in the set of all the interactions $I$ of our system
or if we are more interested in defining a new IS using $(i,j)$.

The previous construction can be repeated with a finite number $i_{1},\ldots,i_{h}$
of simultaneous interactions, so that we can describe arbitrary polyadic
cause-effect relations. Therefore, considering simultaneous  interactions,
we can naturally describe a CS using cause-effect hypergraphs in a
polyadic relationships, see e.g.~\cite{JaBaCr,BaFo21,FBSD21}.

\subsection{\label{subsec:Evolution-equations-1}Evolution equations}

We refer to Sec.~\ref{subsec:Evolution-equations} for the motivations
of the following definitions:
\begin{defn}
\label{def:EE}Let $\mathcal{EI}=(E,\tst,\tend,\mathcal{T},I)$ be
a system of entities and interactions, let $\mathcal{S}=(S,\mathfrak{S},R,x)$
be a system of states for \emph{$\mathcal{EI}$, }and $\mathcal{D}_{i}=(\tist{i},\tong{i},\mathcal{N}_{i})$
the data to run $i$, for each $i\in I$. Therefore
\[
\tfirst(t):=\tfirst:=\inf\left\{ \tarr{i}(t)\mid i\in I,\ \tist{i}(t)=t\right\} \quad\forall t\in[\tst,\tend],
\]
is the \emph{first arrival of all the interactions started at} $t$.
Note that $\tfirst$ is the clock function generated by the values
of $\tarr{i}(t)$ for $i\in I$: in fact, $\tarr{i}$ is piecewise
constant (see Thm.~\ref{thm:startArrNeigh}.\ref{enu:tarrPiecewiseConst}),
and by Def.~\ref{def:dataInt} of $\tarr{i}$, these values are at
most countable. Moreover, $\tarr{i}(t)\ge t=\tist{i}(t)$ by condition
\ref{enu:SA} of Def.~\ref{def:dataInt}.

\noindent In this setting, a \emph{system }$\mathcal{EE}=(\text{\ensuremath{\Delta}},f,\Omega,\mathcal{F},P)$\emph{
for the evolution equations of }$\mathcal{EI}$, $\mathcal{S}$ and
$\left(\mathcal{D}_{i}\right)_{i\in I}$ is given by the following
data which satisfy the following conditions.
\begin{itemize}
\item If the $\tfirst<+\infty$ and $p\in E$, we first define
\[
I_{p}(t):=\{i\in I\mid\tfirst(t)\le\tarr{i}(t)\le\tfirst(t)+\Delta,\ \pa{i}=p\}\quad\forall t\in[\tst,\tend],
\]
where $\Delta\in\R_{\ge0}\cup\{+\infty\}$.
\item Then, we consider the \emph{state of the neighborhood of} $p\in E$
as the function $\stateNeigh_{p}x$ defined by:
\begin{align}
 & \text{If }\exists t'\le t\,\exists i\in I:\ \pa{i}=p,\ t'=\tarr{i}(t'),\ \eps\in\mathcal{N}_{i}(t'),\ \tau\in[t',t]\label{eq:neigh-of-p}\\
 & \text{then }\stateNeigh_{p}x(\tau,\eps):=x_{\eps}(\tau).\nonumber 
\end{align}
\end{itemize}
\noindent These data have to satisfy the following conditions:
\begin{enumerate}
\item \label{enu:probEvolution}$(\Omega_{p},\mathcal{F}_{p},P_{p})$ is
the \emph{probability space for the evolution of $p\in E$} and the
\emph{transition map} \emph{$f_{p}(-,s,\stateNeigh_{p}x_{s})_{3}:\Omega_{p}\ra S_{p}$}
is measurable for all $s\in[\tst,\tend]$.
\end{enumerate}
\begin{enumerate}[resume, leftmargin=*,label=(EE),align=left ]
\item \label{enu:EEm}There exists $\omega\in\Omega_{p}$ such that if
$t\in[\tst,\tend]$, $\tfirst<+\infty$, $I_{p}(t)\ne\emptyset$ and
$\tfirst(t)\le s\le\tfirst(t)+\Delta\le\tend$, then
\begin{equation}
x_{p}(s)=f_{p}\left(\omega,s,\stateNeigh_{p}x_{s}\right),\label{eq:EEm}
\end{equation}
where
\[
\stateNeigh_{p}x_{s}:\tau\in[\tfirst(t),s]\mapsto\stateNeigh_{p}x(\tau,-).
\]
\end{enumerate}
\end{defn}

\noindent Conditions \ref{enu:probEvolution} and \ref{enu:EEm} mathematically
clarify the intuition about the state variable $x_{p}(s)$, which
results as a (possible) stochastic path of our system. In other words:
by running a simulation of the system which follows the algorithm
presented in Sec.~\ref{subsec:Dynamics-of-IS}, we obtain as outcome
a possible value of the state variables $x_{p}(s)$. The label \ref{enu:EEm}
recall \emph{evolution equation}.

To further illustrate these concepts, we can consider the following
simple examples:
\begin{example}
\label{exa:stateChangeFcn}~
\begin{enumerate}[label=\arabic*)]
\item A person $a$ is throwing a stone $p$: we can set the propagator
$r=p$ as the same patient carrying in its resource state the information
of the initial velocity $\vec{v}_{0}$ and position $x_{0}$ (hence
in this example we have $i:a\xra{p,\text{t}}p$ and $\gamma_{i}(t)=(x_{0},\vec{v}_{0})$).
We can also set $\act_{i}^{a}(t)=\act_{i}^{p}(t)=1$, and the starting
time $\tist{i}$ defined by $T^{\text{s}}:=\{\tst\}$, so that $\tist{i}(t)=\tst$
if $t=\tst$ and $\tist{i}(t)=+\infty$ otherwise. The ongoing function
$\tong{i}$ is defined by $T^{\text{o}}:=[\tst,\tend]$ and hence
the arrival function is $\tarr{i}(t)=\tst$ (see also below Sec.~\ref{sec:Classical-models-as-IS}
for an IS defined by an arbitrary ODE). In the time interval $[\tst,\tend]$
the transition function is of the form $f_{p}=f_{p}(s,x_{0},\vec{v}_{0})$
and gives the deterministic dynamics of the stone. In this case we
have a trivial space for stochastic evolution, i.e. $|\Omega_{p}|=1$.
\item In a more ``realistic'' modeling of the same system, we can consider
the initial condition $(x_{0},\vec{v}_{0})$ distributed as a 6-dimensional
normal distribution, so that we have $S_{r}=R_{j}=\R^{6}$ and $\gamma_{j}(t)$
is this normal distribution for another initial interaction with $\tist{j}=\tong{j}=\tarr{j}=\tst$
and corresponding to this random extraction of the initial condition
with normal distributions. The transition function $f_{p}(s,\tst,x_{0},\vec{v}_{0})=x_{0}+\vec{v}_{0}(s-\tst)+\frac{1}{2}\vec{g}(s-\tst)^{2}$
is actually a deterministic function depending on the randomly extracted
initial values $(x_{0},\vec{v}_{0})=(x_{0}(\omega),\vec{v}_{0}(\omega))$
(no additionally randomness is introduced after the stone has been
thrown). The probability space for the evolution of $p$ is hence
again trivial: $|\Omega_{p}|=1$.
\item Let us consider a pedestrian $p$ receiving at time $\tist{i}$ a
signal $r$ from a source $a$, and starting to move in the direction
$\gamma_{i}\in\R^{3}$, $|\gamma_{i}|=1$, with a certain stochastic
deviation, both in the direction and in the magnitude of the velocity.
We will have $f_{p}(s,\omega;\tist{i},\gamma_{i})=x_{0}+\vec{v}(\omega)\cdot(t-\tst{i})$,
where $x_{0}$ is the position of $p$ at time $\tist{i}$ and where
the expected value of $\vec{v}$ in the space $(\Omega_{p},\mathcal{F}_{p},P_{p})$
is $E(\vec{v})=v_{0}\cdot\gamma_{i}$; both $v_{0}$ and $x_{0}$
are taken from the proper state space $S_{p}$ of the patient $p$.
\item Let $i\in I_{p}(t)$ be an interaction starting at $\tist{i}(t)=t$,
arriving at $\tarr{i}(t)\ge t$ and ongoing in the interval $[\tarr{i}(t),t_{k}^{2}]$.
We can have $\tfirst(t)+\Delta>t_{k}^{2}$ if $\Delta$ is small because
of other interactions simultaneously occurring in the interval $[\tfirst(t),\tfirst(t)+\Delta]$,
but, at the same time, we want that $i$ continuously acts on $p$
even after $t':=\tfirst(t)+\Delta$. In order to model this kind of
behaviour in the setting of IS, we clearly have to coherently model
the starting times so that $\tist{i}(t')=t'$, e.g.~using a continuous
time interval for $\tist{i}(-)$.
\item Def.~\ref{def:EE} states minimal conditions satisfied by a large
class of CS. However, it could be also very interesting to consider
IS where at time $\tfirst(t)+\Delta=:t'$ explicitly occur the starting
times of feedback interactions $j:p,n_{1},\ldots,n_{k}\xra{r_{j},\text{fb}}n_{h}$,
where $\{n_{1},\ldots,n_{k}\}\subseteq\mathcal{N}_{i}(t')$ are entities
in the neighborhood of $i$. Since $\tist{j}(t')=t'$ and the transition
function $f_{n_{h}}$ depends also on the state $x_{p}(t'')$ of $p$,
we can say that this is still a particular case of the dynamics described
in \ref{enu:aPatientIsChanged} of Sec.~\ref{subsec:Dynamics-of-IS}.
\end{enumerate}
\end{example}

\subsubsection{\label{subsec:Stochastic-generationClocks}Stochastic generation
of clock functions}

We already specified in Sec.~\ref{subsec:Occurrence-times} and Sec.~\ref{subsec:Clock-functions}
that the clock functions $\tist{i}(t)$, $\tong{i}(t)$ can be thought
of as sample paths generated by model depending distributions. This
can be done using the following procedure:
\begin{enumerate}
\item For each interaction $i\in I$, we consider the \emph{state of the
neighborhood} of $i$ defined as
\begin{align}
 & \text{If }\exists t'\le t:\ t'=\tarr{i}(t'),\ \eps\in\mathcal{N}_{i}(t'),\ \tau\in[t',t]\label{eq:neigh-of-p-i}\\
 & \text{then }\stateNeigh_{i}x(\tau,\eps):=x_{\eps}(\tau).\nonumber 
\end{align}
In general, all the probability distributions of the occurrence times
depend on this neighborhood function, i.e.~on the history of the
interaction $i$.
\item At each time $t$, we have to decide whether $i$ will start after
or at $t$ in a discrete time $t_{j}$ or a continuous time interval
$[t_{k}^{1},t_{k}^{2}]$. Even if, in principle, this can also be
decided randomly, usually it is a model-related choice. 
\item Starting from the previous step and using a model-depending probability
distribution $T_{i}^{\text{s}}(-;\stateNeigh_{i}x)$ on the space
$[t,\tend]$ depending on $\stateNeigh_{i}x$, we can extract either
a sample of the form $t_{j}\in[t,\tend]$ or a pair
\[
(t_{k}^{1},t_{k}^{2})\in\left\{ (t^{1},t^{2})\in[t,\tend]\mid t^{1}<t^{2}\right\} 
\]
(note in both cases the interval $[t,\tend]$). In the first case
we set $\tist{i}(t)=t_{j}$, and in the second one we set $\tist{i}(t')=t'$
for all $t'\in[t_{k}^{1},t_{k}^{2}]$.
\item At each $t$ such that $\tist{i}(t)=t$ (i.e.~$i$ starts at $t$),
using a model-depending probability distribution $T_{i}^{\text{a}}(-;\stateNeigh_{i}x,\tist{i}(t))$
on the space $[\tist{i}(t),\tend]$, depending also on the previous
random value $\tist{i}(t)\in[t,\tend]$, we extract the value of $\tarr{i}(t)$.
Clearly, this could depend on the speed of the propagator $r=\pr{i}$.
We recall that $\tarr{i}(-)$ takes only discrete values and is not
a clock function, see Thm.~\ref{thm:startArrNeigh}.
\item Finally, using a model-depending probability distribution $T_{i}^{\text{o}}(-;\stateNeigh_{i}x,\tarr{i}(t))$
on the space $[\tarr{i}(t),\tend]$, depending also on the previous
random value $\tarr{i}(t)\ge\tist{i}(t)=t$, we extract either if
the interaction $i$ is instantaneously occurring at $t_{j}=\tarr{i}(t)$
or a sample pair $(t_{k}^{1},t_{k}^{2})\in\left\{ (t^{1},t^{2})\in[\tarr{i}(t),\tend]\mid t^{1}<t^{2}\right\} $.
In the first case we set $\tong{i}(t)=t_{j}$, whereas in the second
one we set $\tong{i}(t')=t'$ for all $t'\in[t_{k}^{1},t_{k}^{2}]$.
\item Clearly, by considering trivial probability distributions, the previous
method also includes the deterministic generation of occurrence times.
\end{enumerate}

\subsection{\label{subsec:Interaction-spaces}Interaction spaces}
\begin{defn}
An \emph{interaction space} $\mathfrak{I}=(\mathcal{EI},\mathcal{SA},\mathcal{I},\mathcal{TF})$
is given by considering all the previously defined systems:
\begin{enumerate}
\item A system of entities and interactions $\mathcal{EI}=(E,\tst,\tend,\mathcal{T},I)$.
\item A system of state spaces $\mathcal{SA}=(S,\mathfrak{S},R,x)$\emph{
}for\emph{ $\mathcal{EI}$.}
\item Data $\mathcal{D}_{i}=(\tist{i},\tong{i},\mathcal{N}_{i})$ to run
each interaction $i\in I$\emph{.}
\item A system $\mathcal{EE}=(\text{\ensuremath{\Delta}},f,\Omega,\mathcal{F},P)$\emph{
}for the evolution equations of\emph{ }$\mathcal{EI}$, $\mathcal{S}$
and $\left(\mathcal{D}_{i}\right)_{i\in I}$.
\end{enumerate}
\end{defn}

\noindent After a first look, one can actually recognize that the
previous definitions essentially represent the introduction of several
mathematical notations, and that the important conditions are only
a few, as it is clarified in tables \ref{tab:sysEntitiesInteractions},
\ref{tab:stateSpacesActivation}, \ref{tab:sysInfo} and \ref{tab:sysTransitionFnctns}.

\begin{table}
\noindent \begin{centering}
\begin{tabular}{|c|c|}
\hline 
Symbol & Meaning\tabularnewline
\hline 
\hline 
$E$ & set of interacting entities\tabularnewline
\hline 
$[\tst,\tend]$ & time interval\tabularnewline
\hline 
$\mathcal{T}$ & set of types of interactions\tabularnewline
\hline 
$I$ & set of interactions $i=(a_{1},\dots,a_{n},r,\alpha,p)$\tabularnewline
\hline 
$E_{i}:=\left\{ a_{1},\dots,a_{n},r,p\right\} $ & interacting entities in the interaction $i$\tabularnewline
\hline 
$\ag{i}:=(a_{1},\ldots,a_{n})$ & agents of $i$\tabularnewline
\hline 
$\pa{i}:=p$ & patient of $i$\tabularnewline
\hline 
$\pr{i}:=r$ & propagator of $i$\tabularnewline
\hline 
\end{tabular}
\par\end{centering}
\caption{\label{tab:sysEntitiesInteractions}Summary of Def.~\ref{def:EntitiesAndInteractions}
of system of entities and interactions $\mathcal{EI}=(E,\tst,\tend,\mathcal{T},I)$.}
\end{table}

\begin{table}
\noindent \begin{centering}
\begin{tabular}{|c|c|c|}
\hline 
Symbol & Meaning & Condition\tabularnewline
\hline 
\hline 
$\act_{i}^{e}(t)$ & activation map & $\act{}_{i}^{e}(t)\in[0,1]$\tabularnewline
\hline 
$R_{i}$ & Resources of $i$ & \tabularnewline
\hline 
$\gamma_{i}(t)$ & goods of the interaction $i$ & $\gamma_{i}(t)\in R_{i}$\tabularnewline
\hline 
 & state variable $x_{e}(t)$ and proper state space $S_{e}$ & $\forall e\in E:\ x_{e}(t)\in S_{e}$\tabularnewline
\hline 
\end{tabular}
\par\end{centering}
\caption{\label{tab:stateSpacesActivation}Summary of Def.~\ref{def:stateSpacesActivation}
of system of state spaces and activation maps $\mathcal{S}=(S,\mathfrak{S},R,x)$.}
\end{table}

\begin{table}
\noindent \begin{centering}
\begin{tabular}{|c|c|c|}
\hline 
Symbol & Meaning & Condition\tabularnewline
\hline 
\hline 
$\tist{i}(t)$ & starting time & \ref{enu:CF}\tabularnewline
\hline 
$\tong{i}(t)$ & ongoing time & \ref{enu:CF}\tabularnewline
\hline 
$\tarr{i}(t)$ & arrival time & \ref{enu:SA}, \ref{enu:CE}\tabularnewline
\hline 
$\mathcal{N}_{i}(t)$ & neighborhood of $i$ at time $t$ & \ref{enu:NE}: $E_{i}^{t}\subseteq\mathcal{N}_{i}(t)\subseteq E_{t}$\tabularnewline
\hline 
\end{tabular}
\par\end{centering}
\caption{\label{tab:sysInfo}Summary of Def.~\ref{def:dataInt} of data to
run an interaction $\mathcal{D}=(\tist{i},\tong{i},\mathcal{N}_{i})$.}
\end{table}

\begin{table}
\noindent \begin{centering}
\begin{tabular}{|c|c|c|}
\hline 
Symbol & Meaning & Condition\tabularnewline
\hline 
\hline 
$\Delta$ & time of evolution & $\Delta\in\R_{\ge0}\cup\{+\infty\}$\tabularnewline
\hline 
$(\Omega_{p},\mathcal{F}_{p},P_{p})$ & probability space for the evolution of $p$ & \tabularnewline
\hline 
$f_{p}(\omega,s,\stateNeigh_{p}x_{s})$ & transition function of $p$ & \ref{enu:EEm}\tabularnewline
\hline 
\end{tabular}
\par\end{centering}
\caption{\label{tab:sysTransitionFnctns}Summary of Def.~\ref{def:EE} of
system for the evolution equations $\mathcal{EE}=(\Delta,f,\Omega,\mathcal{F},P)$.}
\end{table}

\section{\label{sec:Classical-models-as-IS}Classical models for complex systems
as interaction spaces}

We can now explain how classical models for complex system can be
embedded as interaction spaces. Of course, these embeddings are injective:
e.g.~if two CA are equal when viewed as IS, then they are necessarily
equal as CA.

Even if in this section we do not always mathematically prove and
detail the corresponding embeddings, in our opinion it is clear that
the following ways to include these classical models as IS are sufficiently
detailed to allow a reconstruction of the initial model from the corresponding
IS. We also see that the IS structure allows one to consider several
interesting generalizations of these classical models of CS. 

\subsection{Continuous dynamical systems}

Assume that the considered system is described by a system of ODE
$x'(s)=F(s,x(s))\in\R^{n}$ for $s\in[\tst,\tend]$ starting from
a given initial state $x_{0}\in\R^{n}$ at $t=\tst$ (we also recall
that any higher order ODE can always be transformed into an equivalent
system of first order ODE). We can think at an IS having a single
entity $p$ with an initial state $x_{p}(\tst)=x_{0}$. At $t\ge\tst$
the dynamics of this IS must be ruled by an evolution equation faithfully
corresponding to this ODE.

The following is only one possible way of seeing a dynamical system
as an IS, and several other embeddings are possible as well, e.g.~because
a dynamical system does not have intrinsic notions of activations,
goods, neighborhood, etc. However, we will see that these additional
notions naturally inspire interesting generalizations. 

We therefore set $E=\{p\}$, $\tst<\tend\le+\infty$, $\mathcal{T}=\{\text{ds}\}$
which means ``dynamical system'', $I=\{(p,p,\text{ds},p)\}$ i.e.~$p\xra{p,\text{ds}}p$,
$x_{p}(t)\in[0,1]^{I}\times\{0\}^{I}\times\R^{n}$. Since there is
only one interaction, in the following we omit the index $i$. We
set trivial activations and goods: $\act^{p}(t):=1$, $\gamma(t):=0$
for all $t$. Occurrence times: $\tist{}$ defined by $T^{\text{s}}:=\{\tst\}$,
so that $\tist{}(t)=\tst$ if $t=\tst$ and $\tist{}(t)=+\infty$
otherwise; $\tong{}$ defined by $T^{\text{o}}:=[\tst,\tend]$, i.e.~$\tong{}(t)=t$
for all $t$; Therefore, the arrival time is given by $\tarr{}(t)=\tst$
for all $t$. Neighborhood: $\mathcal{N}(t)=\{p\}$. Conditions \ref{enu:CF},
\ref{enu:SA}, \ref{enu:CE} and \ref{enu:NE} trivially hold. For
the evolution equation, we set $\Delta=\tend-\tst$, and we have $\tfirst(t)=\tst$
if $t=\tst$ and $\tfirst(t)=+\infty$ otherwise, so that $I_{p}(\tst)=\{i\}$
and $I_{p}(t)=\emptyset$ for $t>\tst$ because $\tfirst(t)=+\infty\le\tarr{}=\tst\le\tfirst(t)+\Delta=+\infty$
is impossible. If $(\tau,\eps)$ lies in the domain of the neighborhood
function $\stateNeigh_{p}x$ (i.e.~if the conditions $p=\pa{i}$,
$t'=\tarr{i}(t')$, $\eps\in\mathcal{N}_{i}(t')$, $\tau\in[t',t]$
hold for some $t'\le t$ and $i\in I$ (see \eqref{eq:neighFunct}),
then necessarily $t'=\tst$ and $\eps=p$, so that
\[
\stateNeigh_{p}x:\tau\in[\tst,t]\mapsto x_{p}(\tau)\in\R^{n},
\]
where we considered only the nontrivial specific state space $S_{p}:=\R^{n}$.
For a deterministic dynamics, we consider a trivial probability space
$\Omega_{p}=\{0\}$. Finally, the assumptions of \ref{enu:EEm} are
$\tfirst(t)<+\infty$ (so that $t=\tst$) and $\tfirst(t)\le s\le\tfirst(t)+\Delta\le\tend$
(i.e.~$\tst\le s\le\tend$) and the evolution equation must be $x_{p}(s)=f_{p}(s,\stateNeigh_{p}x_{s})$,
where the restricted neighborhood function is $\stateNeigh_{p}x_{s}:\tau\in[\tst,s]\mapsto x_{p}(\tau)\in\R^{n}$,
i.e.~it is $x_{p}(-)|_{[\tst,s]}$. As we already anticipated in
Rem.~\ref{rem:EE}.\ref{enu:EE-dynSys}, if we assume that $x'_{p}(s)=F(s,x_{p}(s))$
for all $s\in[\tst,\tend]$, where $F\in\mathcal{C}^{0}([\tst,\tend]\times U,\R^{n})$
is a continuous function and $U\subseteq\R^{n}$ is an open set, we
can define $f_{p}(s,y):=y(\tst)+\int_{\tst}^{s}F(\tau,y(\tau))\diff\tau$
for all $s\in[\tst,\tend]$ and for all $y\in\mathcal{C}^{0}([\tst,\tend],U)$
to have that this ODE is satisfied if and only if $x_{p}(s)=x_{p}(\tst)+\int_{\tst}^{s}F(\tau,x_{p}(\tau))\diff\tau=f_{p}\left(s,x_{p}(-)|_{[\tst,s]}\right)$.
Moreover, $f_{p}$ uniquely determine $F$ as $F(t,x)=\frac{\diff{}}{\diff s}f_{p}(s,x)|_{s=t}$
and this yields the injective embedding from the data $(F,x_{0})$
describing the continuous dynamical system to this IS starting from
$x_{p}(\tst)=x_{0}$.

\subsection{Discrete dynamical systems}

If the dynamical system is described by a recursive equation of the
form $x(k+1)=F(k,x(k))\in\R^{n}$ for all $k=0,\ldots,N$ and $x(0)=x_{0}\in\R^{n}$,
we can set the IS as above, changing only the evolution equation as
described in \eqref{eq:DDS}, where we can think at $x_{p}(-):\{0,\ldots,N\}\ra\R^{n}$
as an arbitrary function. Therefore, the transition function $f_{p}$
uniquely determine the values $V_{F}:=\left\{ F(k,y)\in\R^{n}\mid k\in\{0,\ldots,N\},\ y\in\R^{n}\right\} $
which define all the possible orbits of the given discrete dynamical
system. Therefore, this gives an embedding of the data $(V_{F},x_{0})$
into this IS.

It is natural to think at generalizations of the form:
\begin{itemize}
\item Initial interaction with a starting entity $s\in E$ which is responsible
for setting the initial condition $x_{0}$.
\item Introducing a non-trivial dynamics of goods in a suitable space of
resources (e.g.~described by another dynamical system) corresponds
to coupled dynamical systems.
\item We can also consider several levels of non-Markovian dynamical systems
taking less trivial occurrence times or neighborhoods. For example,
if $\tist{i}(t)<\tarr{i}(t)$ and a previous interaction $j\in I_{p}(\tarr{i}(t))$
returns the state $x_{p}(t')$ back to a previous value $x_{p}(t'-\tau_{\text{D}})$
at some $t'<\tarr{i}(t)$, then we have a delay dynamical system $x'(s)=F(s,x(s-\tau_{\text{D}}))$,
see e.g.~\cite{Ern}. We can also consider a nontrivial neighborhood
and couple the dynamical system with the past dynamics of another
interacting entity.
\item Considering a nontrivial probability space for the evolution of $p$,
we can also describe as IS any stochastic dynamical system.
\item If the system experiences abrupt changes (i.e.~infinite derivatives),
like in collisions, we can similarly describe it as an IS by taking
as $F$ a generalized smooth function, see e.g.~\cite{Gio-Kun-V19}.
\end{itemize}

\subsection{Synchronous and asynchronous cellular automata}

To embed CA as IS, we clearly set cells with their state space as
interacting entities. Depending on the type of cellular automaton,
we can have either local or global interactions $i$, the latter possibly
acting on only a subset of cells. Even if in classical CA we can set
as always active every cell, in more advanced CA we can think at inserting
the activation $\act_{i}^{e}(t)$ as a state variable of some cell
(see e.g.~\cite{Va-Gi-An08a}). Every local interaction has the same
type of neighborhood, which corresponds to that of the cell in the
CA structure. In every local interaction, agents are all the cells
in the neighborhood, and the patient corresponds to the cell on which
the interaction acts. Global interactions can be seen as having only
one agent that equals the patient on which they act. The dynamics
can be synchronous at times $\tist{i}(t)=\tong{i}(t)=\tarr{i}(t)=\tst+k$
if $\tst+k\le t<\tst+k+1$, $k\in\N$, whereas asynchronous dynamics
corresponds to more general occurrence times. Transition functions
$f_{p}$ correspond to the mathematical functions defining the state
change of the cellular automaton. In more advanced CA, interactions
may also depend on a suitable space of resources and goods (see e.g.~\cite{Va-Gi-An08a}),
on continuous state space with stochastic, non-Markovian or time-dependent
rules. Even if every CA can also be seen as a discrete dynamical system,
the setting as IS we are considering here allows one to preserve also
the other structures of the automaton, such as neighborhoods and asynchronous
dynamics, having in this way an embedding.

\subsection{Agent based models}

For ABM, we refer to the mathematical definition given in \cite{Woo02}.
Although agents naturally correspond to interacting entities of IS,
we have to consider that frequently ABM are identified with the corresponding
implementation in an (object oriented) programming language, and the
corresponding mathematical formalization is not always considered.
In that case, the state space of an agent can also include its behavioral
rules or methods, implemented as computer codes. Since IS is a mathematical
theory, such methods have to be associated to a corresponding mathematical
function, but this is clearly always possible because the semantic
of every programming language always has a corresponding mathematical
theory. More generally, even in case of mathematically formalized
ABM, behavioral rules and methods are exactly transition functions
in the language of IS. The environment itself, see e.g.~\cite{Woo02},
has to be considered as an interacting entity. Neighborhoods are defined
by all the entities (agents or environment) from which the methods
(interactions) take the information they need to operate. Note that,
in case another agent is not contained in this neighborhood, it is
completely hidden for the interactions of the considered agent. Since
an interaction is of the form $i:a_{1},\ldots,a_{n}\xra{r,\alpha}p$,
in general interactions are only of local nature, depending on the
agents $a_{1},\ldots,a_{n},p$, on $r$, $\alpha$, and on the cause-effect
relations with other agents. The dynamics is naturally asynchronous,
depending on the signals $r$ sent between agents: these correspond
to propagator entities whose speed has to be modeled only in particular
cases.

\subsection{Master equation based models}

In \cite{Gio24nM}, we prove that the dynamics of a Markovian IS is
described by a master equation. Any Markov model can be seen as a
particular case of one of these Markovian IS. Therefore, this includes
several models used in synergetics, \cite{Sor06,Sch03,Sch02,Wei00}.
Usually, additional structures such as propagators, starting, arrival
and ongoing times, neighborhoods, etc.~are not used in these descriptions.

\subsection{\label{subsec:Networked-dynamical-systems}Networked dynamical systems}

For this kind of model of CS, see e.g.~\cite{New18}. For all time
$t\in[\tst,\tend]$, let $G_{t}=(V_{t},L_{t})$ be a graph with set
of vertices/nodes $V_{t}$ and set of edges/links $L_{t}$. Every
node $e\in V_{t}$ has a state $x_{e}(t)$ belonging to a space $S_{e}$.
This corresponds to an IS where interacting entities are all the nodes
of the network plus unordered pairs of vertices (here, for simplicity,
we consider only non-directed networks)
\[
E:=\bigcup_{t\in[\tst,\tend]}V_{t}\cup\{\{e_{1},e_{2}\}\mid e_{1},e_{2}\in V_{t}\}.
\]
The network can be easily formalized considering $\{0,1\}$-valued
interactions between nodes and corresponding to the adjacency matrix
$G_{t}$: At each time $t$, we define an interaction $i_{l}:=(l,1,\texttt{matr},l)$
for each pair $l=\{e_{1},e_{2}\}\in E$, with $\ag{i_{l}}=\pa{i_{l}}=l$
(and an abstract/trivial propagator), and where at each time $t$
\[
f_{l}(t)=\begin{cases}
1 & l\in L_{t}\\
0 & \text{otherwise}
\end{cases}
\]
We activate only $V_{t}$, i.e.
\begin{equation}
\act_{i}^{e}(t)=1\iff e\in V_{t}.\label{eq:actNDS}
\end{equation}
Each node interacts only with adjacent nodes (which is hence the neighborhood)
and the transition functions correspond to the functions that update
the state of each node, and can hence come, e.g., from the solution
of suitable differential equations. As it is well known, the update
algorithm can be synchronous or asynchronous, and as such it has to
be implemented as the times of an IS. Similarly, one can embed as
IS networked dynamical systems based on hypergraphs.

The setting \eqref{eq:actNDS} clearly states that the activation
function is trivial in a networked dynamical system. We can hence
state that IS allows one to implement a more detailed cause-effect
structure using the activation function. Also propagators, and hence
the spaces of resources, are not used in this formalization of networked
dynamical systems as IS. On the other hand, we could say that IS can
be seen as networked (stochastic) dynamical systems over a cause-effect
weighted directed network with abstract weights given by propagators.

\subsection{Artificial neural networks}

As in the previous case, interacting entities are the neurons of the
network. The structure of the network and the neighborhoods can be
formalized, in the language of IS, using the adjacency matrix as above.
The state of each neuron includes the values of the input variables,
the bias for each one of these inputs, the activation function, the
property of being a start or an end node. State space of propagators
include the weights of the links. State space of neurons could also
include the property of being active with respect to a change of the
inputs or a change of the weights associated to its input links. The
most important interactions depend on the type of learning and, in
general, have propagators (and hence weights) as patient entities.
In case of supervised or reinforcement learning, pairs of examples
can be seen as stochastic goods of suitable propagators. We can also
have neurons as patients if the learning algorithm changes their activation
functions. The dynamics is in general synchronous. If one is interested
in computation times of activation functions, propagators times can
also be considered.

\subsection{Genetic algorithms}

The population of candidate solutions (phenotypes) with their state
space (genotype) are the interacting entities. Stochastic interactions
are clearly the core of these models. Mutation, crossover, inversion
and selection operators can be easily implemented as fitness depending
interactions of an IS. The algorithm is synchronous, but asynchronous
versions can also be implemented, e.g.~by considering more fitted
populations as single interacting entities that spread out their genetic
code over the entire set of interacting entities. In this case, propagators
can be considered, with their times. In this generalization, the introduction
of suitable neighborhoods is also relevant.

\section{Conclusions and future developments}

The present paper represents only the first necessary starting point
to even imagine a mathematical theory of CS, i.e.~the creation of
a common universal mathematical language. The universality of IS theory
allows one to be sure that sufficiently general mathematical results
have a satisfactorily range of applications for a range of different
modeling frameworks of CS. For theorems already going in this direction,
see \cite{Gio24nM,Gio24cas}. Note that this does not force anyone
to switch to IS from his favorite CS setting. 

A precise mathematical universal language also provides the necessary
setting to try a formalization of concepts such as that of CAS, of
hierarchy of CS, of functors preserving cause-effect relations, etc.,
see \cite{Gio24cas} for a mathematical definition of CAS by following
the idea of Zipf's \emph{principle of least effort}, \cite{Zip49},
and \cite{Gio23} for ideas about applications of these notions to
a new approach to artificial intelligence.

Note that the embedding results we showed are not related in any way
to universal machines: we do not restrict to recursive functions and,
first of all, the embeddings are constructed by considering particular
cases of IS without mentioning what kind of functions they are able
to process. 

On the contrary, we already noted that the universality of IS theory
also includes several interesting generalization of well-known modeling
frameworks for CS. In \cite{Va-Gi-An08a,Va-Gi-An08b,Va-Gi-An14},
we already applied this point of view by considering a strong generalization
of the notion of CA for the practical motivations of creating validated
models of urban growth and vehicular traffic. IS theory actually originated
from these practical models, and from the observation that we actually
were considering a very general setting applicable to a large class
of CS.

\end{document}